\documentclass[10pt]{article}

\usepackage{amsthm,amsmath}
\usepackage{stmaryrd}
\usepackage[utf8]{inputenc}
\usepackage{graphicx} 
\usepackage{multirow} 
\usepackage{array}  
\usepackage{amsmath,amsfonts,amsthm,bm}
\usepackage{subfigure}
\newcommand{\tabitem}{~~\llap{\textbullet}~~}
\newcommand\Tstrut{\rule{0pt}{2.6ex}} 
\newcommand\Bstrut{\rule[-0.9ex]{0pt}{0pt}}

\usepackage[]{graphicx}
\RequirePackage{hyperref}
\usepackage{authblk}

\graphicspath{{img/}{fig3/}}

\usepackage[left=2cm,top=2cm, right=2cm, nohead,nofoot]{geometry}

\usepackage{placeins}
\usepackage[utf8]{inputenc}
\usepackage{url}
\usepackage{hyperref}
\usepackage{color}
\usepackage{array,booktabs}
\usepackage{color}

\usepackage{multirow} 
\usepackage{amsmath,amsfonts,amsthm,bm}
\usepackage{subfigure}

\title{Analysis of the Bitcoin blockchain: \\Socio-economic factors behind the adoption}

\author[$1$]{Francesco Parino}
\author[$2$]{Mariano G. Beir\'o}
\author[$1$]{Laetitia Gauvin}

\affil[$1$]{Data Science Laboratory ISI Foundation, Via Chisola 5, Torino, Italy}
\affil[$2$]{Facultad de Ingenier\'ia, Universidad de Buenos Aires, INTECIN (CONICET)\newline Av. Paseo Col\'on 850, Buenos Aires, Argentina}

\date{}
\begin{document}
\maketitle

\begin{abstract}

As the first decentralized digital currency introduced in 2009 together with the blockchain, Bitcoin offers new opportunities both for developed and developing countries. Bitcoin peer-to-peer transactions
are independent of the banking system, thus facilitating foreign exchanges with low transaction fees such as remittances, with a high degree of anonymity. These opportunities 
together with other key factors led the Bitcoin to become extremely popular and made its price skyrocket during $2017$.

However, while the Bitcoin blockchain attracts a lot of attention, it remains difficult to investigate where this attention comes from, due to the pseudo-anonymity of the system, and consequently to appreciate its social impact. Here we make an attempt to characterize the adoption of the bitcoin blockchain by country. In the first part of the work we show that information about the number of Bitcoin software client downloads, the IP addresses that act as relays for the transactions, and the Internet searches about Bitcoin provide together a coherent picture of the system evolution in different countries.
Using these quantities as a proxy for user adoption, we identified several socio-economic indexes such as the GDP per capita, freedom of trade and the Internet penetration as key variables correlated with the degree of user adoption.
In the second part of the work, we build a network of Bitcoin transactions between countries using the IP addresses of nodes relaying transactions and we develop an augmented version of the gravity model of 
trade in order to identify socio-economic factors linked to the flow of bitcoins between countries. In a nutshell our study provides a new insight on the bitcoin adoption by country and on the potential socio-economic drivers of the international
bitcoin flow.

\end{abstract}


\section{Introduction}

Bitcoin is a digital currency created in 2009 as an alternative to the banking system. Not only it offers a payment mechanism without any centralized control (i.e., by institutions, governments, or banks), but it has also introduced the revolutionary concept of the blockchain.
After a continuous growth during the last years, Bitcoin becomes now a solid reality and a fascinating object of study. The possible future applications of the blockchain and of cryptocurrencies in general appear as very promising, even if this technology is relatively new and at the first stage of its evolution.

Studying the Bitcoin system as the most significant implementation of a blockchain cryptocurrency is an important challenge to understand how this decentralized model is behaving in the real-world.
In fact, recent literature abounds with several lines of research linked to the Bitcoin blockchain. 
A large part of the effort is devoted to the study of the blockchain technology itself, in particular to its development 
 \cite{bohme2015bitcoin, antonopoulos2014mastering, kroll2013economics} and  to its application to other domains \cite{swan2015blockchain}.
Another undeniably important line of research concerns the financial and economic aspect, where one of the main questions is related to the evolution of prices \cite{kristoufek2015main, ciaian2016economics, elbahrawy2017evolutionary, guo2018predicting}, and issues
concerning regulatory institutions and policy \cite{bohme2015bitcoin, evans2014economic}.
From a social point of view, the study of the uptake of the Bitcoin proves to be a challenging task due to the pseudo-anonymity of the system. Digital cryptocurrencies such as Bitcoin can have a significant social impact, as they allow for 
fast transactions at low costs, offering a solution for tips, donations, and micro-payments without the need of a banking system, paving the way for their wide adoption. However, as users can generate as 
many pseudonyms as they want, this impact is difficult to quantify. In the direction of investigating the social impact of Bitcoin, previous studies have used either external data such as the number of Bitcoin client software downloads by country, the amount of each fiat currency involved in bitcoin transactions on exchange \cite{krause2016bitcoin}, and bitcoin transaction data \cite{reid2013analysis,tasca2016evolution}. 
To exploit the transactions bitcoin data, a crucial step is the process of deanonymization that consists in grouping pseudonyms belonging to the same users, this technique serves both as a way to evaluate 
the level of privacy of the bitcoin system \cite{androulaki2013evaluating} and to characterize the type of usage \cite{tasca2016evolution,foley2018sex,lischke2016analyzing}. 

Here we propose to combine both Bitcoin transaction data and external data sources to quantify the Bitcoin adoption by country; underlining the main factors that might represent a motivation or a deterrent for the Bitcoin adoption, 
and we explore how this might have evolved over time given the data we have.
Moreover, with the introduction of specific metrics, we build and model an international Bitcoin flow network, and from this model we extract the socio-economic indexes playing a main role in the dynamic of transactions.

We organize the rest of this paper as follows:
Section~\ref{collection_prep} provides an overview of the datasets that we used and a description of the pre-processing stage. We analyze three different external data sources to investigate how relevant they are as proxies to evaluate the Bitcoin user adoption.
In Section~\ref{Bitcoinadoption} we characterize the Bitcoin adoption per country, underlying the relevance of various socio-economic factors and analyzing the adoption trends.
In Section 4 we use deanonymization heuristics on the Bitcoin transaction ledger to build a transaction network of users to which we assign countries based on the Internet addresses (IPs) of the nodes that relay their transactions. We finally model the international Bitcoin flow network using an augmented version of the gravity model of trade, and we explore the socio-economic indexes that are correlated to these flows.
Section 5 summarizes and discusses our results.

\section{\label{collection_prep}Data collection and pre-processing}
As we intend to investigate the Bitcoin adoption per country, beside the Bitcoin transactional data that can be directly extracted from the Bitcoin blockchain using a block explorer service, we gathered
three additional sources of information.
From the Bitcoin system we extracted: the IP address of the first node that has relayed each transaction (available through the API at \texttt{blockchain.info}~\cite{blockchaininfo}), and the number of downloads for Bitcoin Core, one of the major Bitcoin 
clients. Finally, we used information from Google Trends to quantify the collective attention towards Bitcoin. Some details about these datasets are reported in Table~\ref{allDatabase}.

\begin{table}[h!]
\centering
\label{allDatabase}

\begin{tabular}{llll}
\hline
Data          & Coverage period            & Characteristic                                                                                                                                                                      & Source          \Tstrut\Bstrut\\

\hline
Blockchain    & \begin{tabular}[t]{@{}l@{}}2009/01/09 -\\ 2016/02/25\end{tabular}     & \begin{tabular}[t]{@{}l@{}}For each transaction:\\ input addresses, output addresses,\\ amounts, fee, timestamp* \end{tabular}                                                       & \texttt{blockchain.info} \Tstrut\\
IP            & \begin{tabular}[t]{@{}l@{}}2009/01/09 - \\ 2016/02/25**\end{tabular} & \begin{tabular}[t]{@{}l@{}}For each transaction: IP address of \\the first node that relays the transaction.\\ Time resolution of the \\block creation ($\approx10$ minutes)\end{tabular}                                                                                & \texttt{blockchain.info} \Tstrut\\ 

\begin{tabular}[t]{@{}l@{}} Bitcoin client \\ (Bitcoin Core \cite{bitcoincore}) \end{tabular} & \begin{tabular}[t]{@{}l@{}}2009/01/09 -\\ 2015/06/05**\end{tabular}   & \begin{tabular}[t]{@{}l@{}}Number of Bitcoin Core downloads \\ per country, daily aggregated\end{tabular}                                                                             & \texttt{sourceforge.net} \Tstrut\\

\begin{tabular}[t]{@{}l@{}} Google Trends\\ time series \end{tabular} & \begin{tabular}[t]{@{}l@{}}2009/01/09 -\\ 2017/02/01\end{tabular}     & \begin{tabular}[t]{@{}l@{}}Evolution of the number of queries, \\ per country with a week resolution. \\ The data are normalized for each\\ country between 0 and 100.\end{tabular} & Google          \Tstrut\\

\begin{tabular}[!h]{@{}l@{}} Google Trends\\ country interest \end{tabular}  & \begin{tabular}[t]{@{}l@{}}2009/01/09 -\\ 2017/02/01\end{tabular}     & \begin{tabular}[t]{@{}l@{}} Assigns a score to countries \\ based on relative in-country queries. \\ The data are normalized \\  between 0 and 100.\end{tabular} & Google          \Tstrut\Bstrut\\
\hline

\end{tabular}
\raggedright *timestamp of the block creation \Tstrut
\newline
\raggedright **effective coverage period shorter

\caption{Summary of datasets gathered from different sources}

\end{table}

\subsection*{Bitcoin blockchain}
The full Bitcoin blockchain database is freely accessible from the Internet; we collected the list of bitcoin transactions using the API from \texttt{blockchain.info} \cite{blockchaininfo} over a period extended from 2009-01-09 to 2016-02-25\footnotemark\footnotetext{I.e., up to the block of \textit{height} 400000.}. 

In order to send and receive bitcoin, users need to create Bitcoin addresses. For each transaction we gathered the input and output Bitcoin addresses of the users involved as well as the amounts transferred, the fees, the block height and the position relative to the block.
Some general information about the Bitcoin blockchain dataset we collected is reported in Table~\ref{General Info}.
We have used as timestamp for each transaction the Unix timestamp of the creation of the block in which it is contained. In fact, the blockchain does not provide any time information for the transactions, but 
it contains the timestamp of block creation~\cite{BitcoinProtocol}. Considering that several blocks are mined each hour, the block timestamp is a good proxy for our study. 

Regarding the transaction amounts, we converted them from BTC (Bitcoin currency) to USD, using a daily exchange rate, as the Bitcoin price has drastically changed over the years (see Appendix \ref{Converting Bitcoin to USD}). 

\begin{table}[!hbt]
\centering

   \begin{tabular}{lr}
    \hline
    Number of blocks        & $400\,000$ \Tstrut\\
    Number of transactions     & $111\,793\,127$ \\
    Number of unique Bitcoin addresses     & $128\,894\,781$ \\
    \hline
  \end{tabular}
  \label{General Info}
  
  \caption{General statistics about the blockchain dataset collected}
\end{table}

\subsection*{Internet Protocol addresses}
To get an insight about users and their geolocation  we consider the IP of the nodes which relay the transactions in the Bitcoin network. Bitcoin indeed uses a \textit{gossip protocol} in which users communicate
their new transactions to all their connected peers across the network and some studies have shown that connecting to a substantial part of the network the first node/IP 
that communicates a transaction is likely to be its creator \cite{kaminsky2011black, koshy2014analysis, biryukov2014deanonymisation}. 
We  thus  downloaded the IP addresses of the first nodes that act as relayers in each transaction from \texttt{blockchain.info}, with the time resolution of the block creation ($\approx10$ minutes).
As our goal is to perform a socio-economic analysis at the country level, we mapped the IPs into their corresponding countries. This process is described in \ref{IP geo-localisation}. 
Moreover, we are aware that some users use TOR in order to increase their anonymity in the network. TOR is an Internet protocol which reroutes connections through a virtual circuit 
so that the IP address is hidden for the rest of the network. During the geo-localization process we thus filtered those transactions relayed by TOR exit nodes (see Appendix \ref{TOR IP}), which represent less than 0.001\% 
over the total number of transactions.

\begin{figure}[!hbt]
\centering
\caption{Map representing the number of new IP addresses appeared in the Bitcoin system by country. The total number of unique IP addresses that ever appeared in the Bitcoin network in the time interval from 2012-09 to 2014-05, for the countries selected for our study. Countries have been selected based on the activity (IP and clients) in Section \ref{Bitcoinadoption}.}
  \label{mondo}
  \smallskip
  \includegraphics[width=0.75\textwidth]{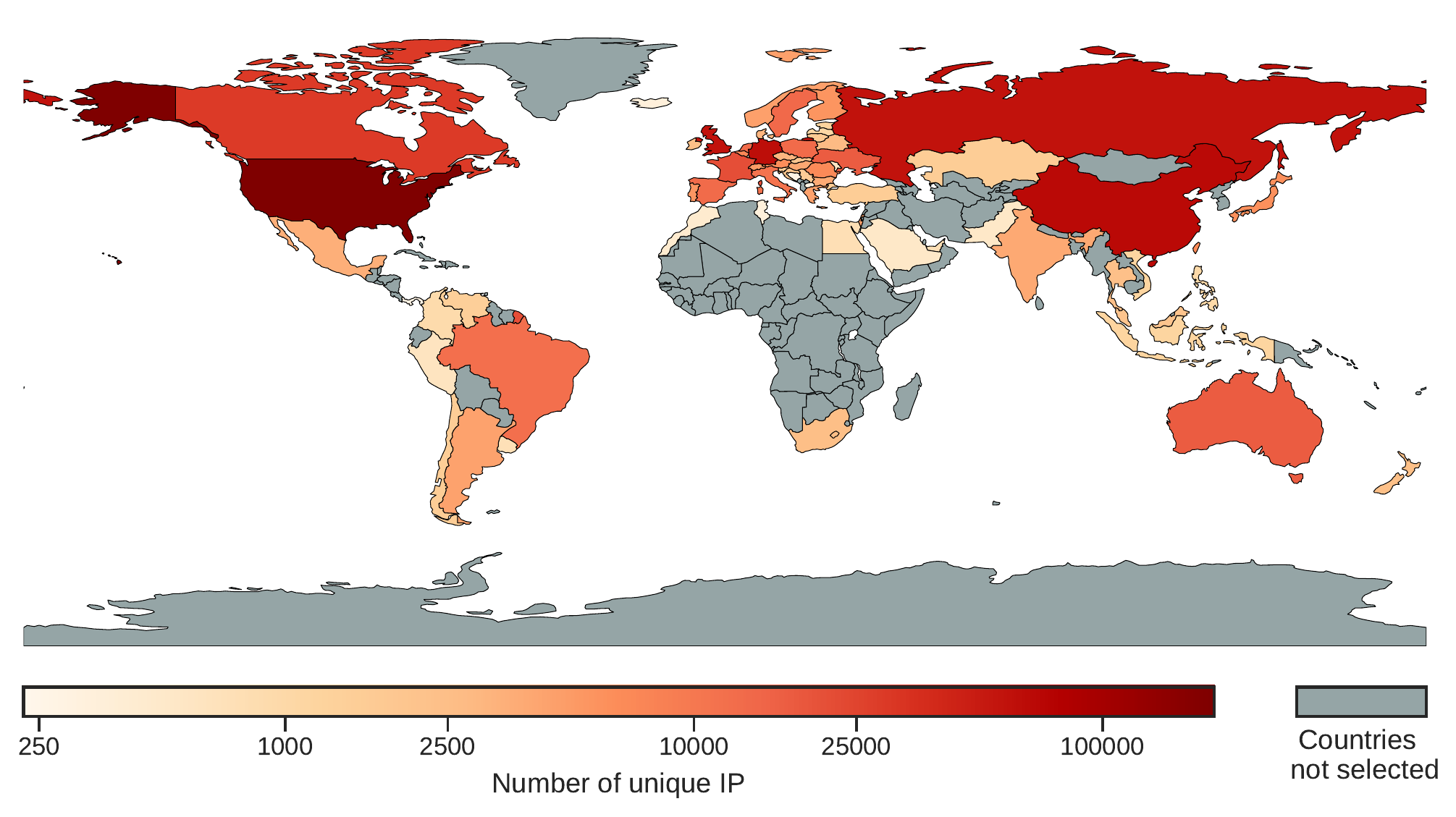}
\end{figure}

One quantity of interest for studying Bitcoin adoption, is the number of such relay node IP adresses that appeared at least one time in the Bitcoin system. Indeed this gives us an idea of the popularity of the Bitcoin
in the different countries as shown in Figure~\ref{mondo} where we reported the number of such IP (each new IP being counted only one time) by countries over our period of study for a selection of countries with enough
activity in the Bitcoin system. In Section \ref{Bitcoinadoption} we explain how we selected these countries.

Looking at the evolution of the  number of new IP appearing in the system (as IP of relay nodes) with time in Figure~\ref{fig:Unique IP and wallet}, we observe a drop in the recorded activity of IP, so we restricted the analysis on the time interval from March 2012 to May 2014.

\subsection*{Bitcoin Client}
\begin{figure}[!hbt]
   \centering
   \caption{Evolution of the number of Bitcoin clients downloaded. World trends, and country-level trends for $2$ of the main countries in term of number of downloads.}
  \label{wallet}
  \smallskip
  \includegraphics[width=.75\textwidth]{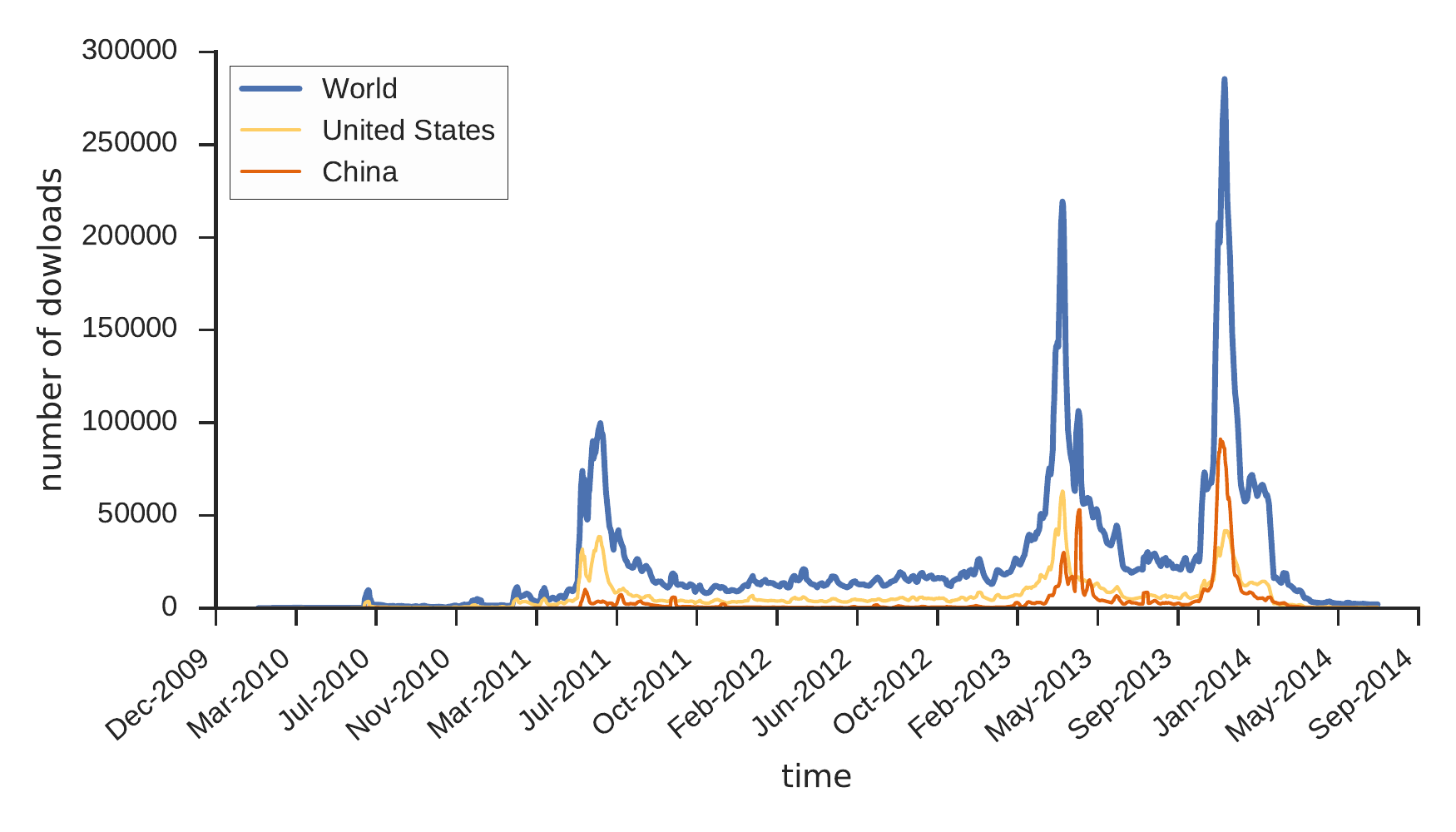}
\end{figure}

To better assess the Bitcoin uptake we also consider the number of Bitcoin Client downloads. 
Generally speaking, a Bitcoin client is a software used to manage and store Bitcoin addresses and make transactions on the Bitcoin network. 
The official Bitcoin client is called \textit{Bitcoin Core}, and it is available from \texttt{sourceforge.net}~\cite{sourceforge}. SourceForge provides some statistics about the downloads, including the total number 
of downloads, daily aggregated by country, as shown in Figure \ref{wallet}.
As other clients exist and some users perform transactions through web-based services, the data from Bitcoin Core does not involve all the Bitcoin users. However, as object of Section \ref{Bitcoinadoption}, we assume that
it gives a reasonable insight on the general distribution and trends of users.
The information saved by the client being reduced in time, as shown in Figure~\ref{fig:Unique IP and wallet}, we limit the interval of analysis on the number of downloads of the client to the period from the 
beginning of 2011 up to May 2014.

\subsection*{Google Trends}
\begin{figure}[!hbt]
\centering
  \caption{Google Trends time series about Bitcoin, for selected countries.}
  \label{google}
  \smallskip
  \includegraphics[width=.75\textwidth]{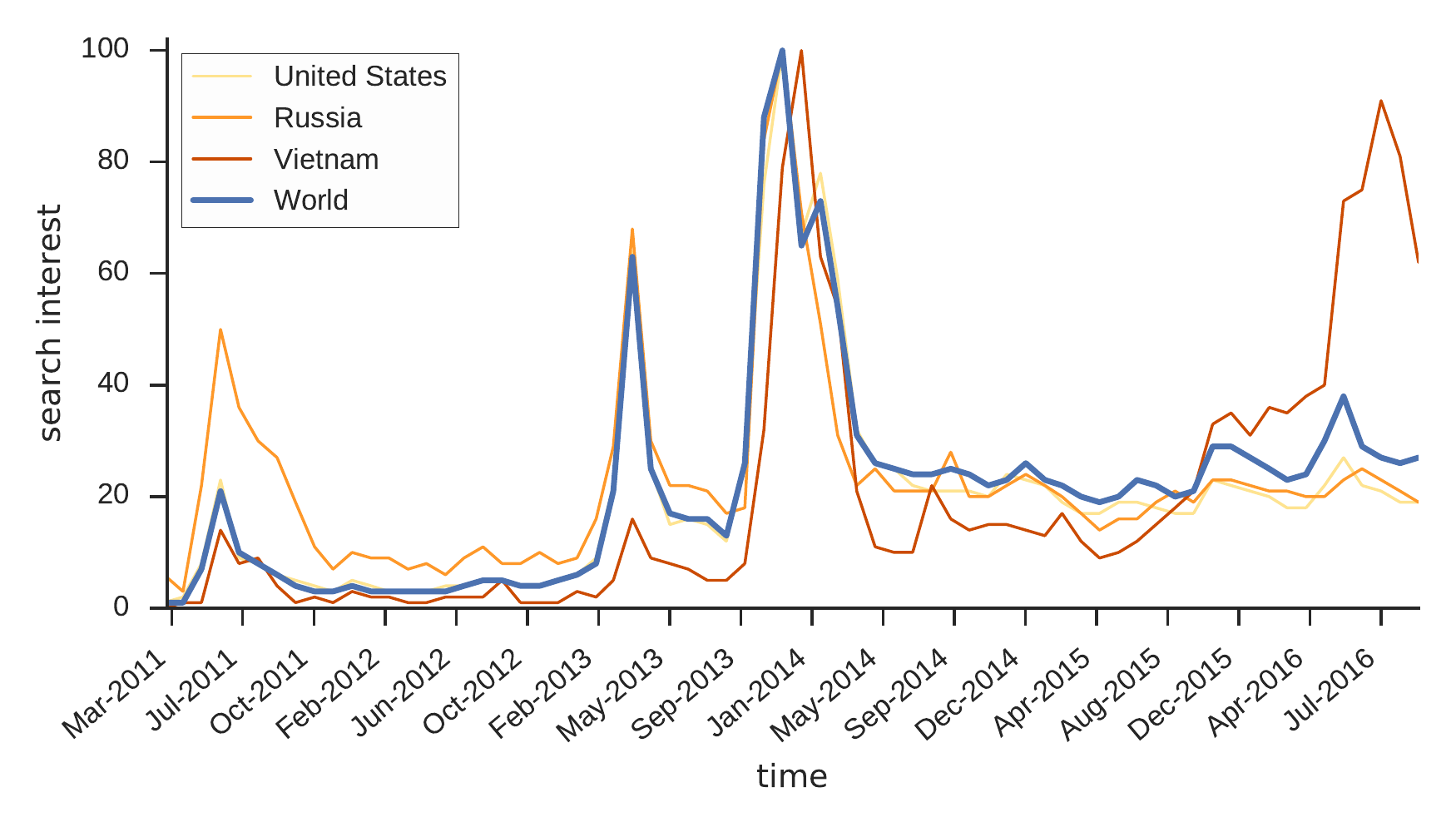}
\end{figure}

Here we use Google Trends as a proxy for the collective attention on the subject, as already proposed in~\cite{puri2016decrypting}.
Figure~\ref{google} provides for each country the evolution of the number of queries relative to the total number of queries done, with a week resolution, for a specific keyword that here we simply set as ``Bitcoin''.
Besides, we extracted the Google's interest by region, using the country's relative number of queries, the scale goes from $0$ to $100$, $100$ being assigned to the country with the highest number of searches on Bitcoin.

\subsection*{Country Socio-Economic indexes}
In order to characterize the adopters of the bitcoin we gathered datasets about socio-economic indexes at the country level with the aim of
exploring the relationship  between these indices and the Bitcoin adoption. We mainly focused on indexes that distinguish the most developed, richest and wealthy countries from the less developed ones. Table~\ref{socioeco} summarizes the indexes that we used.

\begin{table}[]
\centering

\label{socioeco}
\begin{tabular}{l|ll}
\hline
Socio-Economic index & Additional Information                                                                                                          & Source              \Tstrut\Bstrut\\
\hline
Internet Penetration & \begin{tabular}[t]{@{}l@{}} Individuals using the Internet\\ (\% of population) \end{tabular}  & World bank\textsuperscript{[1]}  \Tstrut\\

Population & & World bank          \Tstrut\\

GDP per capita (PPP) & \begin{tabular}[t]{@{}l@{}}GDP per capita converted to \\ international dollars using \\  purchasing power parity rates\end{tabular} & World bank\textsuperscript{[1]}         \Tstrut\\

Inflation & GDP deflator (annual \%) & World bank\textsuperscript{[1]}  \Tstrut\\

HDI & Human develop index & World bank\textsuperscript{[1]} \Tstrut\\

Developing index     & \begin{tabular}[t]{@{}l@{}}Classification:\\ developing/developed\end{tabular}                                                  & United Nations\textsuperscript{[2]}      \Tstrut\\
Developing index     & \begin{tabular}[t]{@{}l@{}}Classification:\\ hight(H)/upper middle(UM)/\\ lower middle(LM)/low(L)\end{tabular}                  & World bank\textsuperscript{[1]}          \Tstrut\\
Over all freedom     & \begin{tabular}[t]{@{}l@{}}Freedom from economical, \\ political and social point of view. \\ Values between 0-100\end{tabular} & Heritage Foundation\textsuperscript{[3]} \Tstrut\\
Freedom to trade     & \begin{tabular}[t]{@{}l@{}}Freedom to trade goods.\\ Values between 0-100\end{tabular}                                          & Heritage Foundation\textsuperscript{[3]} \Tstrut\Bstrut\\
\hline
\end{tabular}

\raggedright [1] https://data.worldbank.org \Tstrut 
\newline
\raggedright [2] http://unctadstat.unctad.org/EN/Index.html 
\newline
\raggedright [3] https://www.heritage.org/index/explore 
\caption{Summary of socio-economic indexes used.}
\end{table}

\section{~\label{Bitcoinadoption}Bitcoin adoption at the country level}
With the goal of appreciating the adoption at the country level, we have identified Bitcoin client downloads, IP of relay nodes and Google Trends as relevant sources of information. Here, we show that these quantities provide a similar and consistent picture of user
and thus, we choose to use them as proxies to study the adoption process.
This preliminary step paves the way for two types of analysis. First, we show how countries with different developing indexes have different trends of adoption and lastly, we explore how country socio-economic indexes are linked to the bitcoin  adoption.

\subsection*{A coherent picture about the users}
The numbers of relay node IP and client downloads are measurements directly related to the blockchain, so that both of them give a direct information of the Bitcoin usage even if none of them can provide a complete picture of 
the users.
In particular, the number of IP addresses does not consider users that do not run a node, and thus do not appear as an IP in the network. On the other side, the number of client downloads provides only information about users
using this specific client. Because of these limitations, we cannot identify the exact number of users per country but a trend of evolution. 
To compare the information given by the numbers of relay node IP and client downloads, we first select countries
 whose activity level permits the analysis. In order to make the selection based on the activity, we computed the medians of the number of client downloads and of the number of different relay nodes IP among all
 countries on moving time windows and we repeatedly filter countries using as thresholds the median of client downloads and the median of the number of unique IP's among all countries. The moving windows are one year wide with a step 
 of one month,
 they cover the period from 2012-03-01 to 2014-05-01. At the end of the filtering process, we select a group of 72 countries,
listed in Table~\ref{Selection of countries}.

On this selection, we explore the relationship between the time series of the numbers of different Ip's and client downloads. We compute the Pearson's correlation coefficient
between the time-series of the number of unique IP appearing in the bitcoin system and the number of bitcoin client downloads both at world wide and country level (time series have been cleaned of the small flucturation by applying a moving window average one month long, with an offset of one
day) the results reported in Table \ref{correlations} indicate high
correlations that confirm that the number of unique IP's and of client downloads give together a coherent picture about the trend of adoption of each country. This supports the point of using both quantities to study the country adoption. 
Additionally, we compute the Spearman's correlation coefficient between the ranking of countries given by IP addresses and client downloads in three different years, arriving to the same conclusion.

\begin{figure}[!hbt]
\centering
	\caption{Summary plots of proxies. Time evolution of the number of Bitcoin client downloads,the number of new IP appearing in the bitcoin system and Google Trends searches on Bitcoin at the worldwide level. 
	The vertical black line marks the limit of database usage.}
  	\label{fig:Unique IP and wallet}
  	\smallskip
  	\includegraphics[width=.8\textwidth]{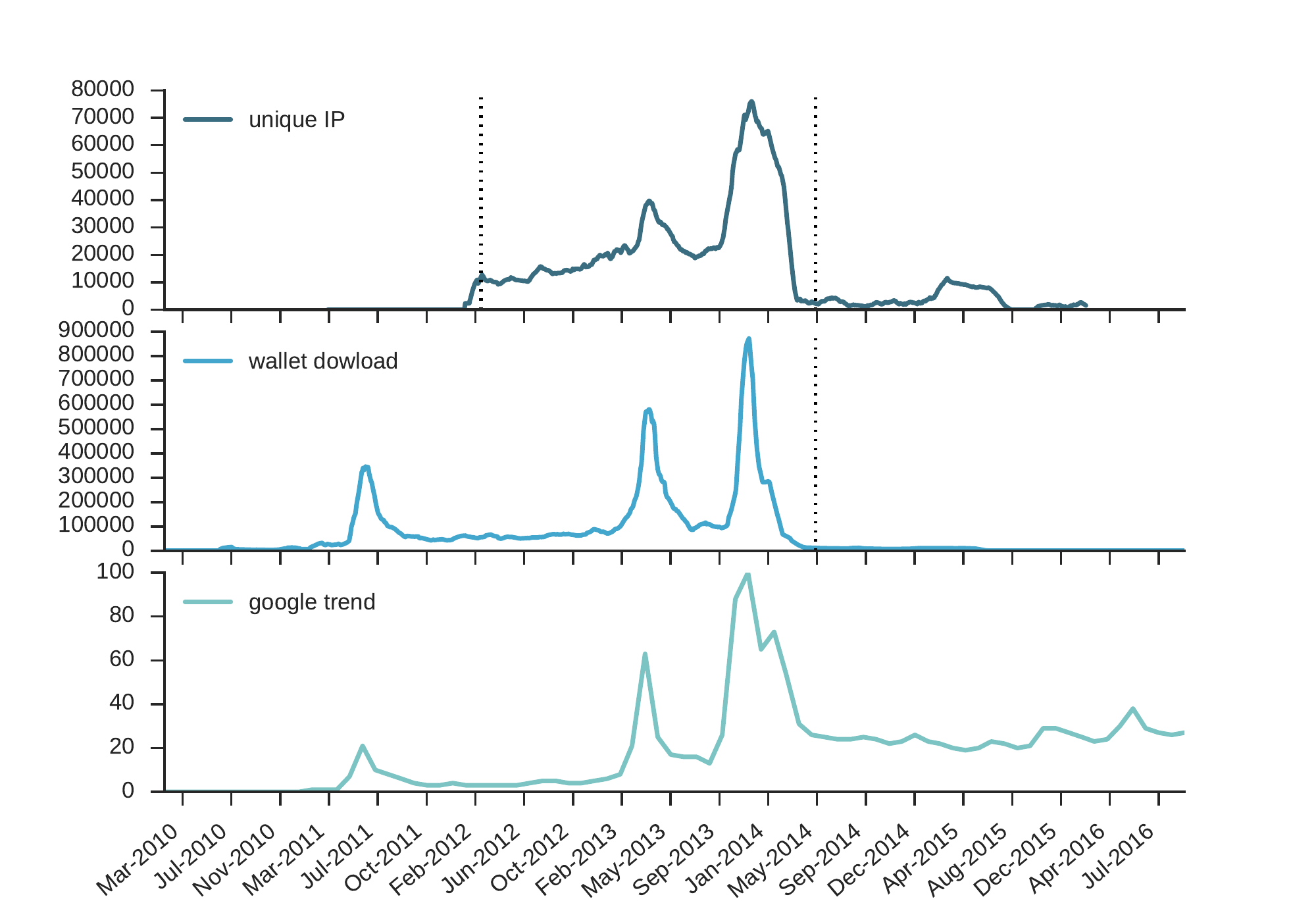}
\end{figure}

We also confronted the Google Trends time series with the numbers of unique IP's and client downloads computing the pairwise Pearson correlations to see if the three data sources give a consistent  picture about
the users.
 Given the high correlations as shown in Table~\ref{correlations}, we assume that Google time series Trends is a well a good indicator of the country Bitcoin adoption, and we suppose that this assumption holds beyond 
 the timespan of the validity
 of IP and client downloads which allows us to discuss long term adoption trends of the selected countries.
 To assess the relevance of the use of Bitcoin search time series for comparing country adoption, we also measured the Spearman correlation between the pairwise rankings of countries by Bitcoin searches, number of Bitcoin client
 downloaded and new IP appearing. Correlations are high apart for the year $2012$ where the signal about Bitcoin searches is too low for allowing comparison between countries. 
 Moreover the country ranking based on Google queries heavily depends on Google's usage by country, which can be very heterogeneous. 
For this reason we wont use the rank provided by Google to extract the socio-economic indices possibly linked to the user adoption, but we can use the country Google time-series to explore the long term trends.

\begin{table}[]
\centering

\label{correlations}
\begin{tabular}{l|ccccc}
\hline
& \begin{tabular}[b]{@{}l@{}}World \\ Time Series\end{tabular} & \begin{tabular}[b]{@{}l@{}}Country\\ Time Series \\ (mean and $\sigma$)\end{tabular} & \multicolumn{3}{c}{Ranking of countries}  \Tstrut\Bstrut\\

\hline

&&& 2012     & 2013		& 2014    \Tstrut\\
IP-Client downloads     & 0.78*		& 0.67 (0.13)	& 0.96     & 0.96    & 0.91    \Tstrut\\
IP-Google     & 0.86~		& 0.76 (0.06)	& 0.04     & 0.85    & 0.79    \\
Client downloads-Google & 0.78~		& 0.61 (0.07)	& 0.08     & 0.87    & 0.74    \Bstrut\\

\hline\Bstrut

\end{tabular}

\raggedright *World level as sum of activities of countries selected in Section~\ref{Bitcoinadoption}

\caption{Correlation between Google Trends time series, number of unique IP's and Bitcoin client downloads. Here we report correlations between the time-series at world level, and the average correlation at country level, during the period from March 2012 to May 2014. Moreover, selecting a period of one year we compute the Spearman's correlation between the countries, ranked using the three proxies.}
\end{table}

\subsection*{Adoption trends: developing versus developed countries}
Using the data from Google Trends we studied the evolution of the collective attention by country from $2009$ to early $2017$. 
As we are interested in the long term trends, we smoothed the Bitcoin search time series by country using a digital low-pass filter to focus on variation on a time scale
of $3$ years.
To study the main trends present in the time series, we built a matrix $A \in \mathbb{R}^{n \times m}$ (where $n$ represents the number of countries and $m$ is the number of points in the time-series),
and we approximated it through non-negative matrix factorization into a product of matrices $W \cdot H$ with $W \in \mathbb{R}^{n \times k}$ and $H \in \mathbb{R}^{k \times m}$. Applying
such appoximation, each country Bitcoin search time series can be represented as a linear combination of $k$ components, stored as the rows
of matrix $H$, and with the coefficients stored in $W$. The number of components has been chosen to be $k=4$ using the bi-cross validation method \cite{owen2009bi}.
 
\begin{figure}
\centering
\caption{Output of the factorization of the matrix representing the Bitcoin search time series. \textit{(Left) Reconstruction error with respect to the number of iterations. \textit{(Right)} Representation of the principal components with $k=4$. The \textit{growing component} (red) represents the trend of the new adopters. The other 3 components are the fluctuating ones, and mostly represent what we call the early adopters. }}
\subfigure{ \includegraphics[width=0.45\textwidth]{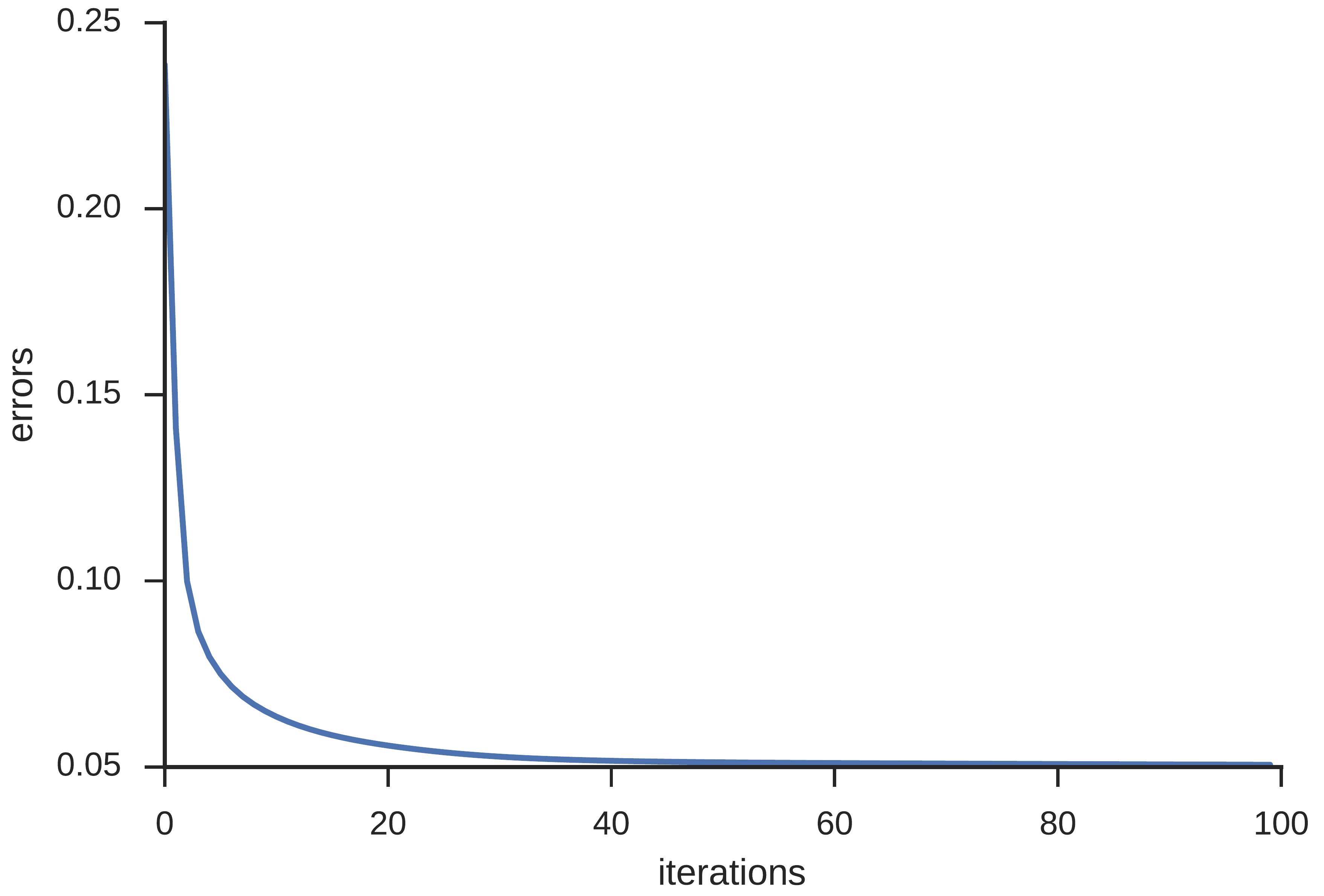}}
\hfill
\subfigure{\includegraphics[width=0.49\textwidth]{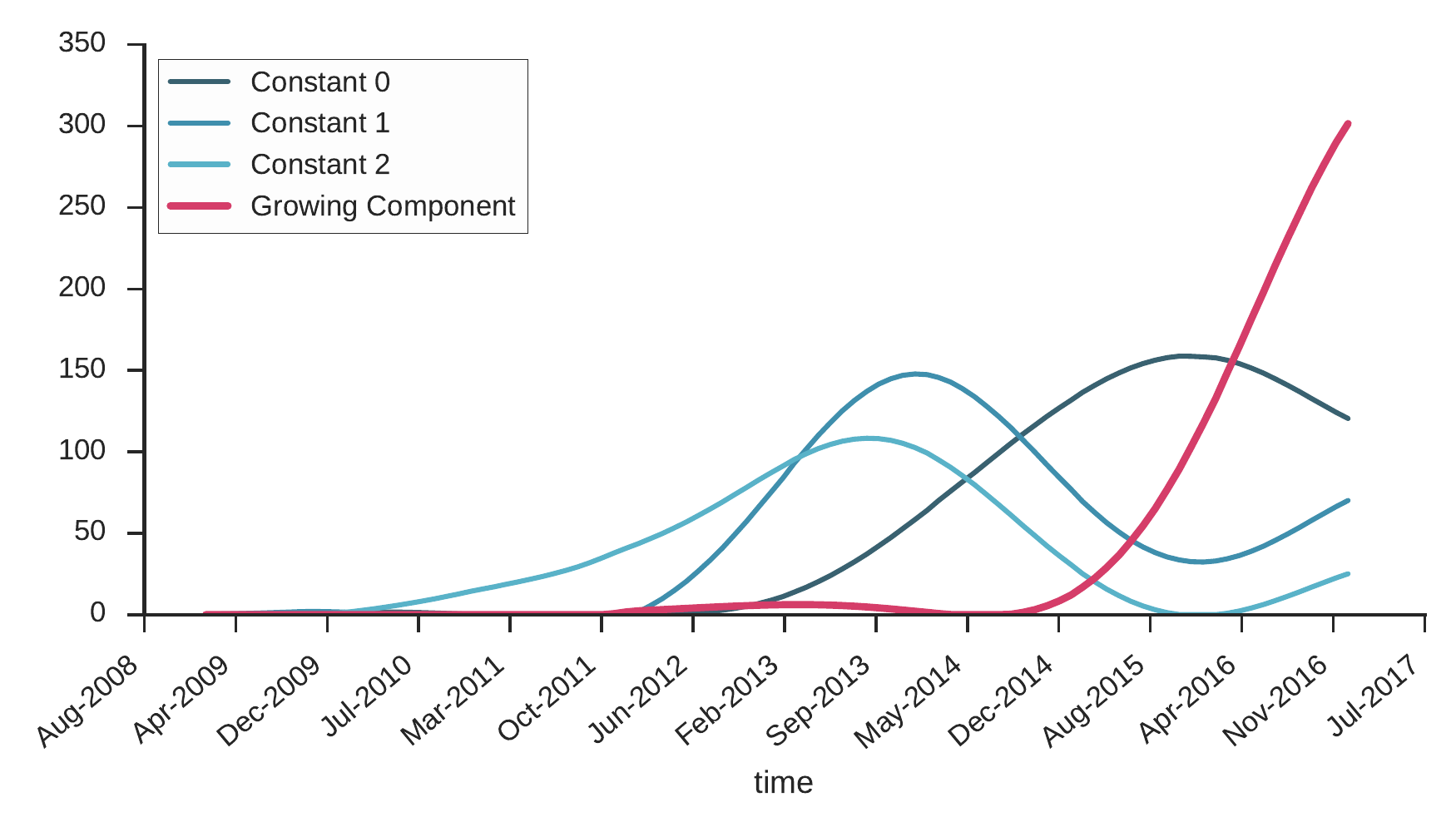}}

\label{ConnectedCC}
\end{figure}

\begin{figure}[!hbt]
\centering
    \caption{Google Trends time series, original, filtered and reconstructed for $6$ countries. For each country we plot: the raw values from Google Trends (green), the filtered trend (red), and the reconstructed trend after the NMF (blue). In the first row we show 2 examples of high developed countries (H) with fluctuating trends. In the second and third row we report  examples of countries with upper middle (UM) and lower middle (LM) development index and a growing trend of adoption.}
    \label{components}
	\smallskip
    \label{abstractPlot}
    \includegraphics[width=0.9\textwidth]{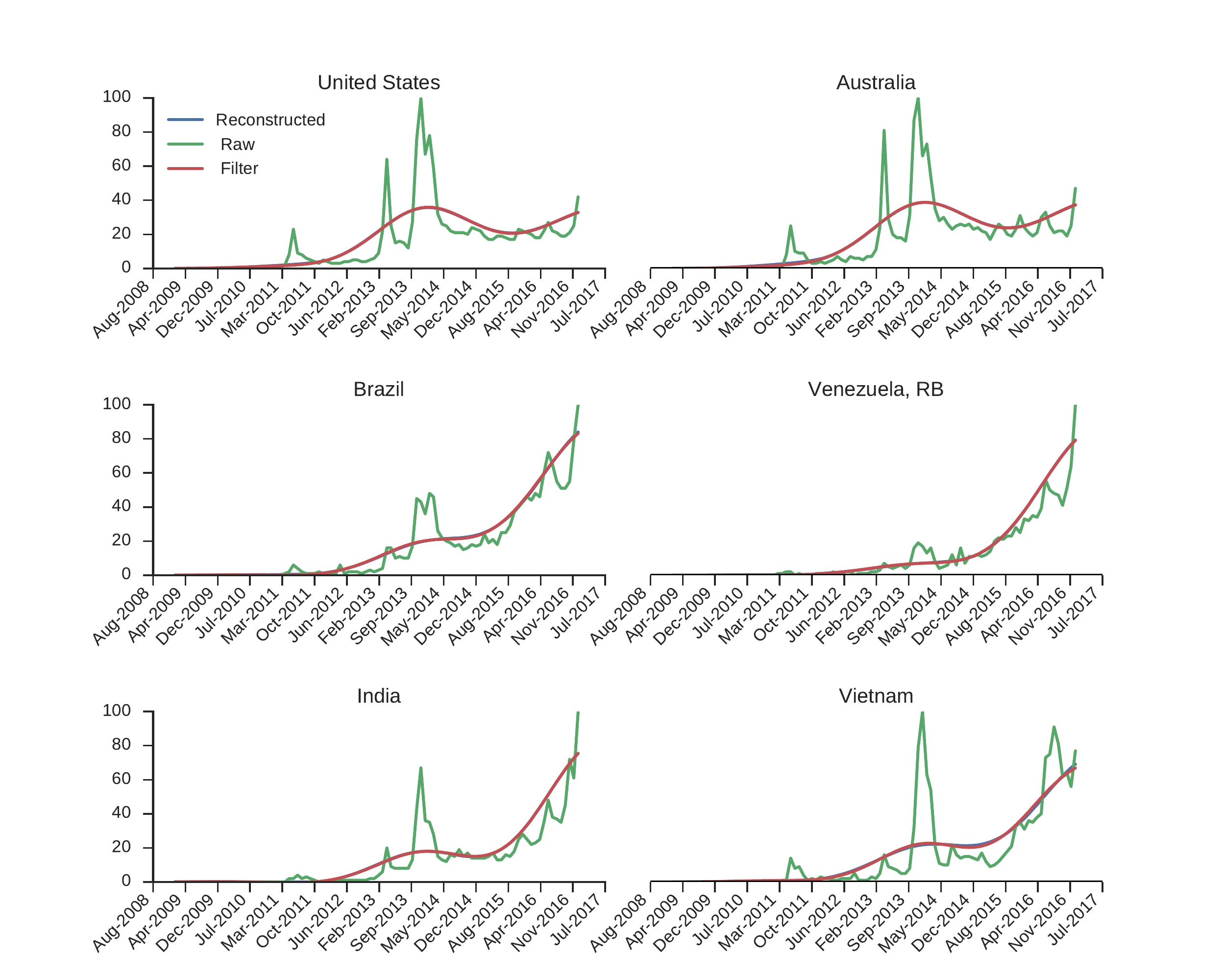}
\end{figure}

In Figure~\ref{components} we show the approximated trends for --smoothed -- time series of $6$ countries and the shape of the $4$ principal components is shown in Figure~\ref{ConnectedCC}: we identified
three components that fluctuate over time, 
and one component that has a clear increasing trend starting from the middle of $2015$.
Looking at the coefficient matrix, $W$, we separated the countries in $2$ groups, those having their highest coefficient for the clear increasing component, we consider them as the \textit{new adopters} of Bitcoin,
and the others whose the main components composing their time series are fluctuating.
As shown in Table \ref{percentage}, grouping countries by development indexes we observed that most of the developed countries are among the early adopters of the Bitcoin, i.e.  the attention was notable already in the early
years of Bitcoin. 
On the other hand, a big part of the developing countries show a recent high interest in Bitcoin. 
\begin{table}[h]
\centering
\label{percentage}

\begin{tabular}{r|cc}
\hline
           & Fluctuating  & Growing \Tstrut\Bstrut\\
\hline
H          & 93\%     & 7\%     \Tstrut\\
LM         & 50\%     & 50\%    \\
UM         & 50\%     & 50\%    \\
Developed  & 95\%     & 5\%     \\
Developing & 46\%     & 54\%    \Bstrut\\
\hline
\end{tabular}
\caption{Repartition of countries between those having mainly an increasing trend and the others for different developing index categories. }
\end{table}

\subsection*{Socio-economic factors behind the adoption}
As measured by the socio-economic indexes, the countries we are analyzing are very heterogeneous, here we attempt to link the different socio-economic indexes with the different trends of adoption.
Focusing on a time interval of one year, we compute the Spearman's correlation coefficient between the rank of countries according to the number of client downloads or number of unique IP addresses (normalized by population)
and the ranking according to different socio-economic indexes.
In the results, reported in Table~\ref{corrSE}, we observe a high positive correlation both with the Internet penetration, GDP per capita (PPP), HDI, and a scarce negative correlation with inflation.
The general picture that emerges is that socio-economic welfare --as present in most developed countries-- has stimulated the Bitcoin adoption, at least for the years 2011, 2012, 2013 and 2014 for which we could carry out this analysis.\\
Beside some expected correlation, like the one regarding the internet penetration that represents an essential condition to participate in the Bitcoin network, the results obtained for the overall freedom and trade freedom are specially interesting.
The two indexes provide a measure of the economic freedom, in particular trade freedom measure the presence of barriers that affect imports and exports of goods and services, and the overall economic freedom index, takes a comprehensive view on the country’s interactions with the rest of the world and the economic and finance policies within the country. 
The good correlation measured suggests that, also if Bitcoin was born with the intention of break obstacles in the way people can exchange money, the general picture about the Bitcoin adoption reveals that the presence 
of policies that promote economical freedom represent a fundamental element in favour of the Bitcoin adoption.

\begin{table}
\centering
\newcolumntype{C}{>{\centering\arraybackslash}p{4em}}

\begin{tabular}{ll|CC|CC}
\hline
Economic index & year & Client downloads  &    p-value & IP &  p-value \Tstrut\Bstrut\\
\hline

GDP per Capita & 2011 &   0.675 &  8.25e-11 &   - &  - \Tstrut\\
    & 2012 &   0.638 &  1.69e-09 &   0.606 &  1.70e-08 \\
    & 2013 &   0.719 &  1.18e-12 &   0.704 &  5.13e-12 \\
    & 2014 &   0.686 &  2.92e-11 &   0.696 &  1.13e-11 \\
    \hline

HDI & 2011 &   0.806 &  1.23e-17 &   - &  - \Tstrut\\
    & 2012 &   0.777 &  1.04e-15 &   0.733 &  2.46e-13 \\
    & 2013 &   0.791 &  1.38e-16 &   0.796 &  6.77e-17 \\
    & 2014 &   0.767 &  3.76e-15 &   0.751 &  3.00e-14 \\
    \hline

Inflation & 2011 &  -0.409 &  3.62e-04 &  - &  - \Tstrut\\
    & 2012 &  -0.223 &  6.00e-02 &  -0.203 &  8.70e-02 \\
    & 2013 &  -0.317 &  6.60e-03 &  -0.277 &  1.83e-02 \\
    & 2014 &  -0.275 &  1.94e-02 &  -0.313 &  7.38e-03 \\
    \hline

Internet Penetration & 2011 &   0.780 &  6.67e-16 &   - &  - \Tstrut\\
    & 2012 &   0.748 &  4.51e-14 &   0.706 &  4.48e-12 \\
    & 2013 &   0.799 &  3.87e-17 &   0.794 &  8.56e-17 \\
    & 2014 &   0.780 &  7.27e-16 &   0.765 &  5.37e-15 \\
    \hline

Overall freedom & 2011 &  0.706 &  5.26e-13 &  - &  - \Tstrut\\
                 & 2012 &  0.678 &  9.08e-12 &  0.639 &  3.19e-10 \\
                 & 2013 &  0.718 &  1.43e-13 &  0.677 &  9.86e-12 \\
                 & 2014 &  0.693 &  2.05e-12 &  0.659 &  5.33e-11 \\
\hline

Trade freedom & 2011 &  0.814 &  1.41e-19 &  - &  - \Tstrut\\
                 & 2012 &  0.817 &  7.85e-20 &  0.789 &  9.51e-18 \\
                 & 2013 &  0.864 &  2.26e-24 &  0.850 &  8.10e-23 \\
                 & 2014 &  0.839 &  9.26e-22 &  0.834 &  2.35e-21 \\
\hline
\end{tabular}

\raggedright - IP data are not available for 2011  \Tstrut
\caption{\label{corrSE} Spearman's correlations between the ranks of countries obtained using the number of unique IP addresses and each socio-economic index collected, 
and the ranks obtained using the number of Bitcoin client downloads and each socio economic index collected}
\end{table}

\section{International Bitcoin flow network}
More than considering the drivers behind the country adoption, in this second section, we attempt to identify the key socio-economic indexes related to the international Bitcoin flow.
The process that leads to the estimation of the Bitcoin flow network consists first of all in a clustering of Bitcoin addresses into users, through a deanonymization process, and a mapping that assigns users to countries.

\subsection*{Identification of users - clustering of addresses}
Bitcoin transactions are based on the utilization of Bitcoin addresses, that are the result of applying a \textit{hashing function} to some input string. Moreover users can create new Bitcoin addresses without limitation in order to hold, receive and send bitcoins; this is computationally cheap and has no cost for them. This anonymizes the users' activities, as we cannot know a priori which users are involved in a transaction, neither which set of Bitcoin addresses belongs to the same user.

However, a partial deanonymization method exists and it permits to reveal the group of Bitcoin addresses likely owned by a single user. This method is based on two heuristics that take inspiration from the underlying functioning of the Bitcoin transaction system~\cite{koshy2014analysis, nick2015data, meiklejohn2013fistful, neudeckercould, doll2014btctrackr, remy2017tracking}. In particular, we base our work starting from the definitions reported in \textit{``Characterizing Payments Among Men with No Names''}~\cite{meiklejohn2013fistful}.
Satoshi Nakamoto, the creator of the Bitcoin system, suggests in his original paper the first heuristic that deals with input addresses~\cite{nakamoto2008bitcoin}. It is based on the fact that the sum of the bitcoins hold in the input addresses of a transaction must be entirely spent and sent to the output address. As a consequence, a user that hold more than one Bitcoin address can provide a certain number of input addresses in order to reach the desired amount he wants to spend. Due to this functioning, the same user might hold all the input addresses of a transaction.
Calling $t$ a transaction and $input(t)$ the set of all the input addresses, we summarize the first heuristic as:

\smallskip
\smallskip
\small
\begin{tabular}{l}
HEURISTIC 1: \Tstrut\Bstrut\\
\begin{tabular}[t]{@{}l@{}}If two (or more) Bitcoin addresses are inputs to the same transaction,\\ they are controlled by the same user. \end{tabular}\\

\hline
\tabitem For a transaction $t$ all $input(t)$ are controlled by the same user \Tstrut\Bstrut\\
\hline
\end{tabular}
\normalsize
\smallskip
\smallskip\\

On the other hand, the second heuristic uses the definition of \textit{shadow addresses}. As described before, the sum of the bitcoins contained in the input addresses has to be entirely spent. As a consequence, the
fraction of the amount that exceeds the value that the sender wants to spend is usually sent to a new Bitcoin address. The latter is called shadow address and is created by the sender to collect back the change. The assumption is then that one of the output addresses can be the shadow address. 

Calling $A_i$ a Bitcoin address we focus on the set of output addresses $\left \{ A_i \right \}_{i \in [\![ 1,n ]\!] }$ of a transaction $output(t)$. We call the number of times the address $A_i$ is used as output of a
transaction as $n^o_{A_i}$.
Focusing on transactions that have at least $2$ output addresses, $n \ge 2$ and adopt the following procedure to identify the shadow addresses:

\smallskip
\smallskip
\small
\begin{tabular}{l}
HEURISTIC 2 \\
\begin{tabular}[c]{@{}l@{}} The shadow address $A_i \in outputs(t)$, if it exists, is controlled by the same user\\ that controls the $inputs(t)$. The definition that brings to the identification of the\\ shadow addresses is:\end{tabular} \Bstrut\\

\hline

\tabitem \begin{tabular}[t]{@{}l@{}} 
$n^o_{A_i} = 1 \text{  and  }  \forall j \in [\![ 1,n]\!] \setminus i \quad  n^o_{A_j}  \neq 1$\\
The Bitcoin address $A_i$ appears only one time as output of a transaction,\\ and there is no other output addresses $A_j$ that satisfies the same condition.
\end{tabular}\Tstrut\\

\tabitem \begin{tabular}[t]{@{}l@{}}
$\forall i \in [\![ 1,n]\!] \quad A_i \notin \text{input(t)}$\\
There is not an explicit self shadow addresses, in the sense that  there are \\ no Bitcoin address that is present both as an input and output of the same transaction. 
\end{tabular}\Bstrut\\

\hline
\end{tabular}
\normalsize
\smallskip
\smallskip\\

After applying the two heuristics, we do not have directly cluster of users, but we only have a partial aggregation at the transaction level. For instance let us assume transactions involving the addresses A, B, C, D, E that are
result in three groups {A, B, C\}, \{A, D\}, and \{D, E\} after the deanonylization process. Then \{A, B, C, D, E\} should be seen as the same user's Bitcoin addresses.
This process of grouping, that can seem straightforward, turns out to be a challenging process considering that the number of incomplete groups scales with the number of transaction considered.
We solve this problem building a network in which Bitcoin addresses represent the nodes and they are linked together if they belong to the same partial group. Then to merge all the incomplete groups, we  extracted the \textit{connected components} of the network. Each connected component represents the complete group of all the user's addresses.

The whole deanonymization process is highly sensitive to mistakes made in the utilization of heuristics. There is the possibility that for some transactions, the principles on which are defined the heuristics are not valid leading to a wrong grouping of Bitcoin addresses.  
This could lead to collapse Bitcoin addresses of different users into a single entity, with the risk of creating users that seem to control a huge number of Bitcoin addresses.
Being aware of this problem, we tried to use the safest heuristics possible, even at the expense of discarding some true linking between Bitcoin addresses.
As some false linking could anyway occurs, the timespan we use for the deanonymization, starts to play a key role; bigger is the period of analysis, bigger is the probability that errors can cause the appearance 
of big clusters of Bitcoin addresses.
Reducing the interval of the analysis might lead to the identification of a large number of small groups of addresses, in other terms the same user might still be splitted in several group of addresses. 

The result shown in this section are based on a deanonymization process that takes into account all the transactions occurred in the year $2013$ (i.e the only year for which we have complete IP information). 
In order to be confident that the results obtained do not heavily depend on the timespan considered for the deanonymization, we carried out the whole modeling analysis -- that is described below --  applying 
deanonymization on different time intervals. 
In particular, we used the period between block 1 and block 400000 (the last in our database), and the one between block $180000$ and $300000$ (that corresponds to the period for which we have the IP information). In both the cases the results are similar and lead to identification of the same socio-economic factors that can explain the international Bitcoin flow.

Finally after running the deanonymization, we can build the transaction network between users, identifying in the transactions the shadow addresses.

\subsection*{Country Association}
Thanks to the deanonymization procedure we can identify transactions in which a specific user appears as sender or creator. Assuming that the first node/IP that relays a transaction is its creator we can associate to each user the list of IP used to send bitcoins.
Using the IP geo-localization here we describe how we associate countries to users.
 A quick look at the user's IP addresses, reveals that we are far from an ideal situation in which every user operates with a single IP, that furthermore is not used by anyone else. Bitcoin services (i.e., the infrastructures that allow users to transact without being a node of the bitcoin network) partially creates this problem as users are seen as using the IP address that belongs to the service. Moreover, a user who does not use services might also use several IP addresses.
To balance the presence of services in the IP addresses usage we build metric, that has the same form of the TFIDF (\textit{term frequency-inverse document frequency}) metric~\cite{Manning:2008:IIR:1394399} commonly used to reflect the importance of words inside documents. This metric respects three main principles that we consider as crucial for the discrimination:
\begin{enumerate}
  \item The metric assigns a score to all possible user's countries, instead of assigning a score to each IP address.
  \item The score rewards the IP usages that are close to the ideal situation, in which an IP address is used just by an user that uses only that IP address.
  \item Being aware that users can use different IP addresses, this metric takes into consideration the ratio between user IP usage and the overall user activity (measured as number of IP addresses ).
\end{enumerate}

The formula used to geo-localize the users is reported in Appendix~\ref{metric}, together with an alternative version, based on similar principles, created to test the robustness of the assignment. As the metric uses the IP information, due to the time limitations shown in Figure \ref{fig:Unique IP and wallet} we carry out this analysis for the restricted timespan from March 2012 to May 2014. The geo-localization process leads to the identification of  destination and origin for 79\% of the transactions in 2013.

In order to test the robustness of the assignment of countries, we compare the result of the 2 versions of the metric, finding that 98\% of users received the same association.
One of the misclassified users is a very active user in $2013$, the TFIDF based method classify it as from United States and the other metric as German. This results in differences in the international flow, but as United States and Germany are both developed countries with similar socio-economic indexes, this will not change the interpretation of the results in the modeling part.

\subsection*{Flow network}
After assigning a country to each user, we created the Bitcoin trade network, in which the nodes represent countries and the weighted links represent the amount of Bitcoins exchanged converted in dollars.
From now on, we will focus on transactions achieved in $2013$ and work with the restricted group of countries analyzed in the first part of the work.
In Figure~\ref{vis} a visualization of the international Bitcoin flow network is displayed.

\begin{figure}[!bth]
	\centering
	\caption{Visualization of the international Bitcoin flow for 2013.  The size of each ribbon is proportional to the amount of Bitcoin expressed in dollars exchanged between 2 countries (the colour of a ribbon identifies the sender country). On the external circle we show the repartition of the flow  in term of sending (external bar) and receiving (external bar) for each country (or group of countries). The groups 1 to 3 have been done by ranking countries by decreasing size and putting together the ones with similar amounts. The representation is done using Circos \cite{Krzywinski18062009}.}
	\label{vis}
  	\smallskip
	\includegraphics[width=.9\textwidth]{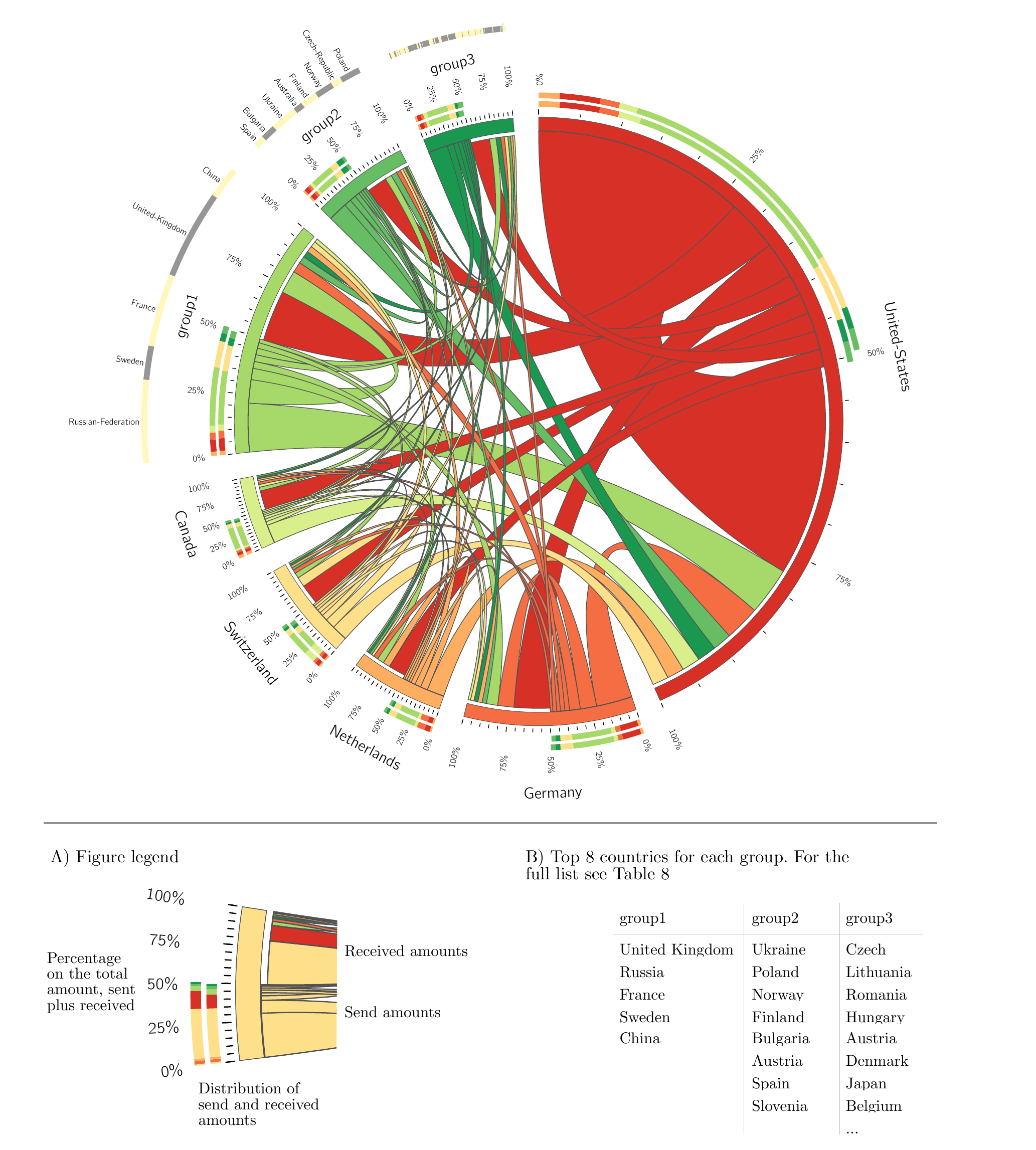}
\end{figure}

\subsection*{Flow modeling}
To understand which socio-economic indexes are potentially explanatory of the Bitcoin flow, we build a model using as a starting point the \textit{gravity model}, introduced by Jan 
Tinbergen in 1962~\cite{tinbergen1962shaping} used to model the bilateral trade flows of different goods and services between countries. The basic form of the model is similar to Newton's law 
of gravitation: it uses socio-economic indexes that represent the economic mass of the country $a$, $M_a$, and which make the interactions stronger, and a variable representing distance between countries, $D_{ab}$,
which decreases the strength of the interactions. Adding a constant $G$, this model takes the form:

\begin{equation}
F_{ab}=G \frac{M_a^{\beta_1}M_b^{\beta_2}}{D_{ab}^{\beta_3}}
\label{gravityF}
\end{equation}
where $F_{ab}$ represents the flow between countries $a$ and $b$ and $\beta_1,\beta_2, \beta_3$ are coefficients that take real values. The traditional approach for fitting the model consists in taking logarithms of both sides, 
leading to a log-log model 
in which it is possible to perform a linear regression~\cite{goldberger1968interpretation} (constant $G$ becomes $\beta_0$). 
\begin{equation}
\ln (F_{ab}) =  \beta_1 \ln (M_a) + \beta_2 \ln (M_b) - \beta_3 \ln (D_{ab}) + \beta_0
\end{equation}
Here we use an \textit{augmented gravity model} \cite{bergstrand1985gravity, lewer2008gravity, carrere2006revisiting}, which means we are considering additional variables. 
Calling ${\{X_i^{ab}\}}_{i \in \llbracket 1,n \rrbracket}$, the $n$ variables that might be either single country quantities (e.g. the masses $M_a$ and $M_b$) or quantities related to the 
couple of countries $(a,b)$
 (e.g. the distance $D_{ab}$), the model can now be written as: 
\begin{equation}
\ln (F_{ab}) = \sum_{i = 1}^{n} {\beta_{i} \: \ln{X_i^{ab}}}  + \beta_0  
\end{equation}
Positive $\beta_i$ are associated to variables $X_i^{ab}$ that contribute to the mass of countries while negative values instead represent variables that act like distances.
However, this approach cannot model the zero observations, and the estimation of the log-linearized equation by least squares (OLS) can lead to significant biases under heteroskedasticity~\cite{silva2006log}. As an alternative, it is possible to work with its multiplicative form, as shown in Equation~\ref{linlog}, replacing the linear regression by a Poisson regression.
\begin{equation}
F_{ab}=\exp\left \{        \sum_{i = 1}^{n} {\beta_{i} \: \ln{X_i^{ab}}}  + \beta_0             \right \}
\label{linlog}
\end{equation}

The vector $\boldsymbol{\beta} = [\beta_0 \dots \beta_n]$ is estimated maximizing the likelihood :
\begin{equation}
l\left( {\left. \boldsymbol{\beta}  \right|X,F} \right) = \sum\limits_{\forall (a,b)} {\left( {{F^{ab}}\cdot \left( {\boldsymbol{\beta}  \cdot {\mathbf{x^{ab}}}} \right) - {e^{\boldsymbol{\beta} \cdot \mathbf{x^{ab}}}}} \right)} 
\end{equation}
where $F$ is a vector containing the Bitcoin flows between $m$ pairs of countries and $X$ is an $m \times (n+1)$ matrix, where each column is given by a vector $\mathbf{x^{ab}}$ whose the values are 
the variables ${X_i^{ab}}_{i \in \llbracket 1,n \rrbracket}$ 
concatenated to a $1$ that is introduced to take into account the constant term $\beta_0$. 

Here we use the following group of variables frequently encountered in the literature on trade: population, distance, GDP per capita, and interaction variables that identify countries with a common language or
geographic border. 
Besides, we consider  Freedom to Trade, Overall Freedom, and Internet Penetration, as we observed (see Table~\ref{corrSE}) that they are linked to the Bitcoin adoption.
Additionally to the datasets described before, we downloaded datasets containing information about countries that share a geographic border or the language~\cite{mayer2011notes}. Finally,
we used a database that reports the distance between each pair of countries, measured using city-level data to account for the geographic distribution of population inside each nation~\cite{mayer2011notes}\footnotemark\footnotetext{In this last dataset we miss information about Serbia, so we did not consider it in the model.}.

As a preprocessing step, the variables are standardized, and the Bitcoin flow is estimated in millions of dollars. 
We then model the flow network maximizing the likelihood introduced below with all the variables mentioned.
Despite the heterogeneity of countries in term of trends of adoption, the model achieves a $R^2$ score of $0.68$. 
This confirms that the socio-economic indexes taken into consideration are good indicators for the international Bitcoin flow.

In order to identify the main drivers of the Bitcoin flow among these socio-economic indexes, we perform a variable selection. To this aim, we introduce $L_{1}$ regularization to the model. In practice we estimate the variables which minimize
\begin{equation}
-l\left ( \beta |  X,Y \right ) + \lambda \sum_{i=1}^{n}\left | \beta_i \right |
\end{equation}
where  $\lambda$ controls the importance of the regularization term. We repeat this process increasing the value of $\lambda$ from $10^{-3}$ to  $10^{1}$. This leads to the cancellation of the coefficients of the variables that contributes less to the flow.
Here we use a 10-folds cross validation in order to set the value of $\lambda$, and we use the average \textit{mean squared error} over the different folds as metric to compare the model's performance.
Each of the $10$ folds is related to a list of pairs of countries chosen at random. We use as test set the \textit{k}th fold that contains $m_k$ couples of countries $(ab)_k$. Calling $f_{\lambda}^{-k}$ the model
with the regularization term $\lambda$ trained excluding the \textit{k}th fold, we compute the cross-validation error $CV_k$ as the mean squared error on the test set:
\begin{equation}
CV_{k}(\lambda)=\frac{1}{m_k}\sum_{\forall (ab)_k}\left (   F^{ab_k}  - f_{\lambda}^{-k} \left ( \mathbf{x}^{ab_k} \right ) \right ) ^2
\end{equation}
Then, we compute the mean of $CV_{k}(\lambda)$, the  standard deviation ($SD$) and the standard error ($SE$) as: 
\begin{equation}
SD(\lambda) = \sqrt{\text{var} \left (  CV_1(\lambda)...CV_k(\lambda)    \right ) } \\
SE(\lambda) =  \frac{SD(\lambda)}{\sqrt{k}} 
\end{equation}
In Figure \ref{CV} for each value of $\lambda$ tested we show the mean squared error.
\begin{figure}[!bth]
	\centering
	\caption{Cross-validation error. The average mean squared error (MSE) for different values of $\lambda$. The right vertical line represent the value of $\lambda$ selected with the one standard error rule, and the left one indicate the position of the minimum}
	\label{CV}
  	\smallskip
	\includegraphics[width=.70\textwidth]{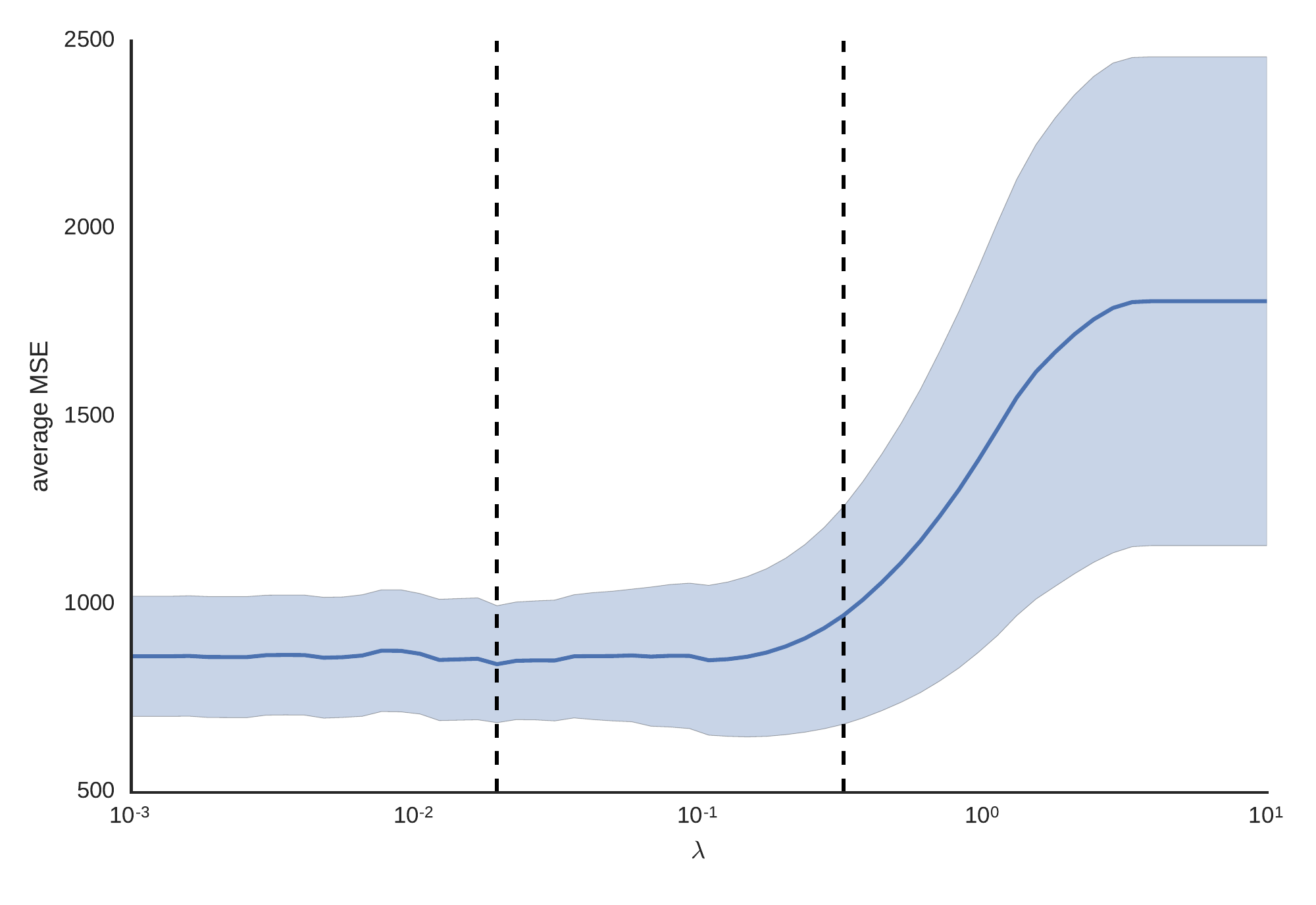}
\end{figure}
As the fluctuations of the cross validation error are small on a large range of $\lambda$ values, instead of choosing the model with the value $\lambda_{min}$ that minimizes the error, 
we apply the \textit{one standard error rule}. This means that we set $\lambda=\widehat{\lambda}$ where $\widehat{\lambda}$ is such that :
\begin{equation}
CV(\widehat{\lambda}) = CV(\lambda_{min}) + SE(\lambda_{min})    
\end{equation}
Fitting the flow with the model described with $\lambda=\widehat{\lambda}= 0.3$, we identify the main variables (among all those selected for the study) that are explanatory of the flow, 
the coefficients we found for those are reported in Table~\ref{poissoncoeff}. In that case the adjustred $R^2$ is equal to $0.57$ even though some variables have been dropped.
\begin{table}[!h]
\centering
\caption{Coefficients for the variables of the augmented gravity model with regularization, applied to the Bitcoin flow. Positive coefficients make stronger the interaction between countries increasing the Bitcoin flow, the negative ones act like a distance decreasing
the interaction and the bitcoins exchanged. Overall freedom index, common border and common language are excluded as their respective coefficients drop to $0$ during the variable selection made through the regularizatoin.}
\begin{tabular}{ll}
\hline
Variables             & Coefficient  \Tstrut\Bstrut\\
\hline
Population In            & 1.65   \Tstrut\\
Population Out           & 1.62   \\
Distance                 & -0.31  \\
Trade Freedom In         & 0.89   \\
Trade Freedom Out        & 0.88   \\
Internet Penetration In  & 0.21   \\
Internet Penetration Out & 0.24   \\
GDP per capita In        & 1.02   \\
GDP per capita Out       & 0.98   \Bstrut\\
\hline
Constant                 & -1.31  \Tstrut\Bstrut\\
\hline
\end{tabular}
\label{poissoncoeff}
\end{table}
On one hand, the coefficient of the overall economic freedom index drops to $0$ due the variable selection meaning that even though this index takes a comprehensive view of the economic freedom of a country it turns out not to be a key 
factor to describe the flow. On the other hand, 
the more specific trade freedom index appears to be, after population, one of the most important variable to describe the flow.
The geographic distance appears as an impediment for the flow. In a nutshell, internet penetration, trade freedom together with GDP and population reveal to be the main potential drivers for the Bitcoin flow.

\newpage
\section{Conclusion}

The blockchain infrastructure offered by cryptocurrencies like Bitcoin is attracting a variety of areas such as trade, finances, government and policy. However, it reveals a challenging task to quantify this attraction and the 
adoption by countries.

In this work we aimed at understanding which are the main factors that pushed the adoption of the Bitcoin as the first blockchain technology in many countries. In order to do this, we applied different techniques for deanonymizing and geolocating the users. Due to the partial anonymity offered by the blockchain, discovering the location of the Bitcoin users is a challenging task; we tackled this problem by combining a series of proxies with the transactional data coming from the Bitcoin public ledger.
In the first part of the work we showed that the
number of IP addresses associated to the relay nodes of the transactions, the number of Bitcoin client downloads,
and the interest measured by Google Trends, all give a coherent picture
about user adoption by country, even though each of them provides only a partial view of the Bitcoin system. Relying on this result, we analyzed the Bitcoin search time series
to explore the evolution of the country attention, and we observed the presence of 
a net increasing trend of attention from 2015 to 2017, coming mostly from developing countries.
Besides, considering the Bitcoin client downloads and IP addresses as proxies for user adoption, we have seen that the adoption is highly correlated with the population, 
the GDP per capita, the freedom of trade and the Internet penetration for the years 2012, 2013 and 2014. 
Overall we also confirm that the Bitcoin adoption trends have not been homogeneous all around the world: since its introduction, Bitcoin has had a fast growth in many developed countries, while its adoption in developing countries increased very slowly.

In the second part of the work, we focused on the Bitcoin flow that is still little explored in the literature, in particular due to the issues related to deanonymization.
and we observed that freedom of trade, GDP and population appear as key  variables to explain the Bitcoin flow.

While this work gives a hint on the socio-economic indexes linked with the Bitcoin adoption, it relies on to use of the IP addresses of relay nodes, which are available only for a restricted time period. As future work, the exploration of other datasources beside \texttt{blockchain.info} could provide IP information for a different period. Another interesting path to overcome this problem would be to model the behavior observed in the transactions with respect to the current distribution of the IP usage accessible, in order to infer the international Bitcoin flows for longer periods of time.

Though we consider here the total flow generated by users and business services (i.e. web-based services like gambling, exchanging, market, mining, clients, etc.), a separate analysis of these types of flows and activities could also help to understand how the Bitcoin is being currently used.

\newpage
\bibliographystyle{bmc-mathphys}

\newcommand{\BMCxmlcomment}[1]{}

\BMCxmlcomment{

<refgrp>

<bibl id="B1">
  <title><p>Bitcoin: Economics, technology, and governance</p></title>
  <aug>
    <au><snm>B{\"o}hme</snm><fnm>R</fnm></au>
    <au><snm>Christin</snm><fnm>N</fnm></au>
    <au><snm>Edelman</snm><fnm>B</fnm></au>
    <au><snm>Moore</snm><fnm>T</fnm></au>
  </aug>
  <source>Journal of Economic Perspectives</source>
  <pubdate>2015</pubdate>
  <volume>29</volume>
  <issue>2</issue>
  <fpage>213</fpage>
  <lpage>-38</lpage>
</bibl>

<bibl id="B2">
  <title><p>Mastering Bitcoin: unlocking digital cryptocurrencies</p></title>
  <aug>
    <au><snm>Antonopoulos</snm><fnm>AM</fnm></au>
  </aug>
  <publisher>CA, USA: O'Reilly Media, Inc.</publisher>
  <pubdate>2014</pubdate>
</bibl>

<bibl id="B3">
  <title><p>The economics of Bitcoin mining, or Bitcoin in the presence of
  adversaries</p></title>
  <aug>
    <au><snm>Kroll</snm><fnm>JA</fnm></au>
    <au><snm>Davey</snm><fnm>IC</fnm></au>
    <au><snm>Felten</snm><fnm>EW</fnm></au>
  </aug>
  <source>Proceedings of WEIS</source>
  <pubdate>2013</pubdate>
  <volume>2013</volume>
  <fpage>11</fpage>
</bibl>

<bibl id="B4">
  <title><p>Blockchain: Blueprint for a new economy</p></title>
  <aug>
    <au><snm>Swan</snm><fnm>M</fnm></au>
  </aug>
  <publisher>CA, USA: O'Reilly Media, Inc.</publisher>
  <pubdate>2015</pubdate>
</bibl>

<bibl id="B5">
  <title><p>What are the main drivers of the Bitcoin price? Evidence from
  wavelet coherence analysis</p></title>
  <aug>
    <au><snm>Kristoufek</snm><fnm>L</fnm></au>
  </aug>
  <source>PloS one</source>
  <publisher>Public Library of Science</publisher>
  <pubdate>2015</pubdate>
  <volume>10</volume>
  <issue>4</issue>
  <fpage>e0123923</fpage>
</bibl>

<bibl id="B6">
  <title><p>The economics of BitCoin price formation</p></title>
  <aug>
    <au><snm>Ciaian</snm><fnm>P</fnm></au>
    <au><snm>Rajcaniova</snm><fnm>M</fnm></au>
    <au><snm>Kancs</snm><fnm>d</fnm></au>
  </aug>
  <source>Applied Economics</source>
  <publisher>Taylor \& Francis</publisher>
  <pubdate>2016</pubdate>
  <volume>48</volume>
  <issue>19</issue>
  <fpage>1799</fpage>
  <lpage>-1815</lpage>
</bibl>

<bibl id="B7">
  <title><p>Evolutionary dynamics of the cryptocurrency market</p></title>
  <aug>
    <au><snm>ElBahrawy</snm><fnm>A</fnm></au>
    <au><snm>Alessandretti</snm><fnm>L</fnm></au>
    <au><snm>Kandler</snm><fnm>A</fnm></au>
    <au><snm>Pastor Satorras</snm><fnm>R</fnm></au>
    <au><snm>Baronchelli</snm><fnm>A</fnm></au>
  </aug>
  <source>Royal Society open science</source>
  <publisher>The Royal Society</publisher>
  <pubdate>2017</pubdate>
  <volume>4</volume>
  <issue>11</issue>
  <fpage>170623</fpage>
</bibl>

<bibl id="B8">
  <title><p>Predicting short-term Bitcoin price fluctuations from buy and sell
  orders</p></title>
  <aug>
    <au><snm>Guo</snm><fnm>T</fnm></au>
    <au><snm>Antulov Fantulin</snm><fnm>N</fnm></au>
  </aug>
  <source>arXiv preprint arXiv:1802.04065</source>
  <pubdate>2018</pubdate>
</bibl>

<bibl id="B9">
  <title><p>Economic aspects of bitcoin and other decentralized public-ledger
  currency platforms</p></title>
  <aug>
    <au><snm>Evans</snm><fnm>D</fnm></au>
  </aug>
  <pubdate>2014</pubdate>
</bibl>

<bibl id="B10">
  <title><p>Bitcoin: Implications for the developing world</p></title>
  <aug>
    <au><snm>Krause</snm><fnm>M</fnm></au>
  </aug>
  <pubdate>2016</pubdate>
</bibl>

<bibl id="B11">
  <title><p>An analysis of anonymity in the bitcoin system</p></title>
  <aug>
    <au><snm>Reid</snm><fnm>F</fnm></au>
    <au><snm>Harrigan</snm><fnm>M</fnm></au>
  </aug>
  <source>Security and privacy in social networks</source>
  <publisher>Springer</publisher>
  <fpage>197</fpage>
  <lpage>-223</lpage>
</bibl>

<bibl id="B12">
  <title><p>The evolution of the bitcoin economy: extracting and analyzing the
  network of payment relationships</p></title>
  <aug>
    <au><snm>Tasca</snm><fnm>P</fnm></au>
    <au><snm>Liu</snm><fnm>S</fnm></au>
    <au><snm>Hayes</snm><fnm>A</fnm></au>
  </aug>
  <pubdate>2016</pubdate>
</bibl>

<bibl id="B13">
  <title><p>Evaluating user privacy in bitcoin</p></title>
  <aug>
    <au><snm>Androulaki</snm><fnm>E</fnm></au>
    <au><snm>Karame</snm><fnm>GO</fnm></au>
    <au><snm>Roeschlin</snm><fnm>M</fnm></au>
    <au><snm>Scherer</snm><fnm>T</fnm></au>
    <au><snm>Capkun</snm><fnm>S</fnm></au>
  </aug>
  <source>International Conference on Financial Cryptography and Data
  Security</source>
  <pubdate>2013</pubdate>
  <fpage>34</fpage>
  <lpage>-51</lpage>
</bibl>

<bibl id="B14">
  <title><p>Sex, Drugs, and Bitcoin: How Much Illegal Activity is Financed
  Through Cryptocurrencies?</p></title>
  <aug>
    <au><snm>Foley</snm><fnm>S</fnm></au>
    <au><snm>Karlsen</snm><fnm>J</fnm></au>
    <au><snm>Putni{\c{n}}{\v{s}}</snm><fnm>TJ</fnm></au>
  </aug>
  <pubdate>2018</pubdate>
</bibl>

<bibl id="B15">
  <title><p>Analyzing the bitcoin network: The first four years</p></title>
  <aug>
    <au><snm>Lischke</snm><fnm>M</fnm></au>
    <au><snm>Fabian</snm><fnm>B</fnm></au>
  </aug>
  <source>Future Internet</source>
  <publisher>Multidisciplinary Digital Publishing Institute</publisher>
  <pubdate>2016</pubdate>
  <volume>8</volume>
  <issue>1</issue>
  <fpage>7</fpage>
</bibl>

<bibl id="B16">
  <url>https://blockchain.info</url>
</bibl>

<bibl id="B17">
  <url>https://bitcoin.org/en/bitcoin-core/</url>
</bibl>

<bibl id="B18">
  <url>https://en.bitcoin.it/wiki/</url>
</bibl>

<bibl id="B19">
  <title><p>?Black ops of TCP/IP</p></title>
  <aug>
    <au><snm>Kaminsky</snm><fnm>D</fnm></au>
  </aug>
  <source>Black Hat USA</source>
  <pubdate>2011</pubdate>
</bibl>

<bibl id="B20">
  <title><p>An analysis of anonymity in bitcoin using p2p network
  traffic</p></title>
  <aug>
    <au><snm>Koshy</snm><fnm>D</fnm></au>
    <au><snm>Koshy</snm><fnm>P</fnm></au>
    <au><snm>McDaniel</snm><fnm>P</fnm></au>
  </aug>
  <pubdate>2014</pubdate>
</bibl>

<bibl id="B21">
  <title><p>Deanonymisation of clients in Bitcoin P2P network</p></title>
  <aug>
    <au><snm>Biryukov</snm><fnm>A</fnm></au>
    <au><snm>Khovratovich</snm><fnm>D</fnm></au>
    <au><snm>Pustogarov</snm><fnm>I</fnm></au>
  </aug>
  <source>Proceedings of the 2014 ACM SIGSAC Conference on Computer and
  Communications Security</source>
  <pubdate>2014</pubdate>
  <fpage>15</fpage>
  <lpage>-29</lpage>
</bibl>

<bibl id="B22">
  <url>https://sourceforge.net/projects/bitcoin/</url>
</bibl>

<bibl id="B23">
  <title><p>Decrypting Bitcoin Prices and Adoption Rates using Google
  Search</p></title>
  <aug>
    <au><snm>Puri</snm><fnm>V</fnm></au>
  </aug>
  <pubdate>2016</pubdate>
</bibl>

<bibl id="B24">
  <title><p>Bi-cross-validation of the SVD and the nonnegative matrix
  factorization</p></title>
  <aug>
    <au><snm>Owen</snm><fnm>AB</fnm></au>
    <au><snm>Perry</snm><fnm>PO</fnm></au>
  </aug>
  <source>The annals of applied statistics</source>
  <publisher>JSTOR</publisher>
  <pubdate>2009</pubdate>
  <fpage>564</fpage>
  <lpage>-594</lpage>
</bibl>

<bibl id="B25">
  <title><p>Data-Driven De-Anonymization in Bitcoin</p></title>
  <aug>
    <au><snm>Nick</snm><fnm>JD</fnm></au>
  </aug>
  <source>PhD thesis</source>
  <pubdate>2015</pubdate>
</bibl>

<bibl id="B26">
  <title><p>A fistful of bitcoins: characterizing payments among men with no
  names</p></title>
  <aug>
    <au><snm>Meiklejohn</snm><fnm>S</fnm></au>
    <au><snm>Pomarole</snm><fnm>M</fnm></au>
    <au><snm>Jordan</snm><fnm>G</fnm></au>
    <au><snm>Levchenko</snm><fnm>K</fnm></au>
    <au><snm>McCoy</snm><fnm>D</fnm></au>
    <au><snm>Voelker</snm><fnm>GM</fnm></au>
    <au><snm>Savage</snm><fnm>S</fnm></au>
  </aug>
  <source>Proceedings of the 2013 conference on Internet measurement
  conference</source>
  <pubdate>2013</pubdate>
  <fpage>127</fpage>
  <lpage>-140</lpage>
</bibl>

<bibl id="B27">
  <title><p>Could Network Information Facilitate Address Clustering in
  Bitcoin?</p></title>
  <aug>
    <au><snm>Neudecker</snm><fnm>T</fnm></au>
    <au><snm>Hartenstein</snm><fnm>H</fnm></au>
  </aug>
</bibl>

<bibl id="B28">
  <title><p>Btctrackr: finding and displaying clusters in bitcoin</p></title>
  <aug>
    <au><snm>Doll</snm><fnm>A</fnm></au>
    <au><snm>Chagani</snm><fnm>S</fnm></au>
    <au><snm>Kranch</snm><fnm>M</fnm></au>
    <au><snm>Murti</snm><fnm>V</fnm></au>
  </aug>
  <source>Princeton University, USA</source>
  <pubdate>2014</pubdate>
</bibl>

<bibl id="B29">
  <title><p>Tracking bitcoin users activity using community detection on a
  network of weak signals</p></title>
  <aug>
    <au><snm>Remy</snm><fnm>C</fnm></au>
    <au><snm>Rym</snm><fnm>B</fnm></au>
    <au><snm>Matthieu</snm><fnm>L</fnm></au>
  </aug>
  <source>International Workshop on Complex Networks and their
  Applications</source>
  <pubdate>2017</pubdate>
  <fpage>166</fpage>
  <lpage>-177</lpage>
</bibl>

<bibl id="B30">
  <title><p>Bitcoin: A peer-to-peer electronic cash system</p></title>
  <aug>
    <au><snm>Nakamoto</snm><fnm>S</fnm></au>
  </aug>
  <pubdate>2008</pubdate>
</bibl>

<bibl id="B31">
  <title><p>Introduction to Information Retrieval</p></title>
  <aug>
    <au><snm>Manning</snm><fnm>CD</fnm></au>
    <au><snm>Raghavan</snm><fnm>P</fnm></au>
    <au><snm>Sch\"{u}tze</snm><fnm>H</fnm></au>
  </aug>
  <publisher>New York, NY, USA: Cambridge University Press</publisher>
  <pubdate>2008</pubdate>
</bibl>

<bibl id="B32">
  <title><p>Shaping the world economy; suggestions for an international
  economic policy</p></title>
  <aug>
    <au><snm>Tinbergen</snm><fnm>J</fnm></au>
  </aug>
  <source>Books (Jan Tinbergen)</source>
  <publisher>Twentieth Century Fund, New York</publisher>
  <pubdate>1962</pubdate>
</bibl>

<bibl id="B33">
  <title><p>The interpretation and estimation of Cobb-Douglas
  functions</p></title>
  <aug>
    <au><snm>Goldberger</snm><fnm>AS</fnm></au>
  </aug>
  <source>Econometrica: Journal of the Econometric Society</source>
  <publisher>JSTOR</publisher>
  <pubdate>1968</pubdate>
  <fpage>464</fpage>
  <lpage>-472</lpage>
</bibl>

<bibl id="B34">
  <title><p>The gravity equation in international trade: some microeconomic
  foundations and empirical evidence</p></title>
  <aug>
    <au><snm>Bergstrand</snm><fnm>JH</fnm></au>
  </aug>
  <source>The review of economics and statistics</source>
  <publisher>JSTOR</publisher>
  <pubdate>1985</pubdate>
  <fpage>474</fpage>
  <lpage>-481</lpage>
</bibl>

<bibl id="B35">
  <title><p>A gravity model of immigration</p></title>
  <aug>
    <au><snm>Lewer</snm><fnm>JJ</fnm></au>
    <au><snm>Berg</snm><fnm>H</fnm></au>
  </aug>
  <source>Economics letters</source>
  <publisher>Elsevier</publisher>
  <pubdate>2008</pubdate>
  <volume>99</volume>
  <issue>1</issue>
  <fpage>164</fpage>
  <lpage>-167</lpage>
</bibl>

<bibl id="B36">
  <title><p>Revisiting the effects of regional trade agreements on trade flows
  with proper specification of the gravity model</p></title>
  <aug>
    <au><snm>Carrere</snm><fnm>C</fnm></au>
  </aug>
  <source>European Economic Review</source>
  <publisher>Elsevier</publisher>
  <pubdate>2006</pubdate>
  <volume>50</volume>
  <issue>2</issue>
  <fpage>223</fpage>
  <lpage>-247</lpage>
</bibl>

<bibl id="B37">
  <title><p>The log of gravity</p></title>
  <aug>
    <au><snm>Silva</snm><fnm>JS</fnm></au>
    <au><snm>Tenreyro</snm><fnm>S</fnm></au>
  </aug>
  <source>The Review of Economics and statistics</source>
  <publisher>MIT Press</publisher>
  <pubdate>2006</pubdate>
  <volume>88</volume>
  <issue>4</issue>
  <fpage>641</fpage>
  <lpage>-658</lpage>
</bibl>

<bibl id="B38">
  <title><p>Notes on CEPII’s distances measures: The GeoDist
  database</p></title>
  <aug>
    <au><snm>Mayer</snm><fnm>T</fnm></au>
    <au><snm>Zignago</snm><fnm>S</fnm></au>
  </aug>
  <pubdate>2011</pubdate>
</bibl>

<bibl id="B39">
  <url>https://www.torproject.org</url>
</bibl>

<bibl id="B40">
  <title><p>Characterization of tor exit-nodes</p></title>
  <aug>
    <au><snm>Schaap</snm><fnm>A</fnm></au>
  </aug>
  <source>Proc. 18th Twente Student Conf</source>
  <pubdate>2013</pubdate>
</bibl>

<bibl id="B41">
  <title><p>IP geolocation databases: Unreliable?</p></title>
  <aug>
    <au><snm>Poese</snm><fnm>I</fnm></au>
    <au><snm>Uhlig</snm><fnm>S</fnm></au>
    <au><snm>Kaafar</snm><fnm>MA</fnm></au>
    <au><snm>Donnet</snm><fnm>B</fnm></au>
    <au><snm>Gueye</snm><fnm>B</fnm></au>
  </aug>
  <source>ACM SIGCOMM Computer Communication Review</source>
  <publisher>ACM</publisher>
  <pubdate>2011</pubdate>
  <volume>41</volume>
  <issue>2</issue>
  <fpage>53</fpage>
  <lpage>-56</lpage>
</bibl>

<bibl id="B42">
  <title><p>Circos: An information aesthetic for comparative
  genomics</p></title>
  <aug>
    <au><snm>Krzywinski</snm><fnm>MI</fnm></au>
    <au><snm>Schein</snm><fnm>JE</fnm></au>
    <au><snm>Birol</snm><fnm>I</fnm></au>
    <au><snm>Connors</snm><fnm>J</fnm></au>
    <au><snm>Gascoyne</snm><fnm>R</fnm></au>
    <au><snm>Horsman</snm><fnm>D</fnm></au>
    <au><snm>Jones</snm><fnm>SJ</fnm></au>
    <au><snm>Marra</snm><fnm>MA</fnm></au>
  </aug>
  <source>Genome Research</source>
  <pubdate>2009</pubdate>
  <url>http://genome.cshlp.org/content/early/2009/06/15/gr.092759.109.abstract</url>
</bibl>

</refgrp>
} 
 
\newpage
\appendix

\section{Additional information on the datasets}

\subsection{Converting Bitcoin to USD's} \label{Converting Bitcoin to USD}
From 2011 it has been possible to exchange Bitcoin with fiat currencies The law of supply and demand dictates the price.
The value of 1 Bitcoin (BTC) is usually given in U.S dollars (USD) and this value has changed drastically over the last years, from cents to thousands of dollars.
Because of this, considering the exchange amounts directly in BTC's might not be representative of their real value, and thus we converted BCT's into USD's using a daily exchange rate obtained from \texttt{blockchain.info}.

\subsection{TOR IP addresses} \label{TOR IP}
The use of Bitcoin offers a good degree of anonymity through the use of pseudonyms but it does not guarantee a complete privacy. For this reason, some users cover their daily activities using TOR \cite{TOR}.
TOR is an Internet protocol which reroutes its users' connections through a virtual circuit so that the user IP address is hidden for the rest of the network, who sees instead the IP address of the last node used by the TOR protocol, also called \textit{Exit Node} and which belongs to TOR~\cite{schaap2013characterization}.
It is possible to get the full historical list of IPs used as \textit{Exit Nodes} by TOR at \texttt{https://collector.torproject.org}, including their corresponding timespan of activity.
Comparing this list against our dataset we found around $50\,000$ TOR transactions that we removed from our study, as by definition we cannot geo-localise them.

\subsection{IP geo-localisation} \label{IP geo-localisation}
IP addresses can be mapped into countries using several online geolocation tools. These tools rely on different sources of information for building their private databases (probably reversal DNS, pings, and the WHOIS protocol, among others), and their results have been validated at the country level~\cite{poese2011ip}. Particularly, we used the \texttt{http://freegeoip.net} API to map every IP address that appeared as a node in the Bitcoin transaction network into a country. As historical records are not available for mapping the node's country at the moment of each transaction, we used the information available at January 2017. We assume that only a negligible fraction of IP addresses have changed location during these years.

\begin{table}[h]
\centering
\caption{Selected countries for this work. Moreover, as done for the visualization in Figure~\ref{vis} the countries are divided in three groups based on the sum of amount sensed and received.}
\label{Selection of countries}
\begin{tabular}{lllll}
              & Group1             & Group2         & \multicolumn{2}{l}{Group3}                    \Tstrut\Bstrut\\

\hline

United States & France             & Poland         & Colombia             & Slovak \Tstrut\\
Germany       & Russian Federation & Australia      & South Africa         & India                  \\
Netherlands   & United Kingdom     & Ukraine        & Argentina            & Pakistan               \\
Switzerland   & Sweden             & Bulgaria       & Croatia              & Morocco                \\
Canada        & China              & Czech Republic & Venezuela, RB        & Japan                  \\
              &                    & Finland        & Thailand             & Indonesia              \\
              &                    & Spain          & Hong Kong & Malaysia               \\
              &                    & Norway         & Moldova              & Lithuania              \\
              &                    &                & Luxembourg           & Hungary                \\
              &                    &                & Singapore            & Belarus                \\
              &                    &                & Mexico               & Bosnia and Herzegovina \\
              &                    &                & Denmark              & Vietnam                \\
              &                    &                & Latvia               & Portugal               \\
              &                    &                & Slovenia             & Taiwan                 \\
              &                    &                & Uruguay              & Greece                 \\
              &                    &                & Peru                 & Belgium                \\
              &                    &                & Panama               & Romania                \\
              &                    &                & New Zealand          & Philippines            \\
              &                    &                & Iceland              & Kazakhstan             \\
              &                    &                & Israel               & Chile                  \\
              &                    &                & Turkey               & Brazil                 \\
              &                    &                & Ireland              & Tunisia                \\
              &                    &                & Arab Emirates & Egypt, Arab Rep.       \\
              &                    &                & Italy                & Korea, Rep.            \\
              &                    &                & Estonia              & Austria                \\
              &                    &                & Malta                & Saudi Arabia           \\
              &                    &                & Macedonia, FYR       &                       
\end{tabular}
\end{table}

\section{Methodology}
\subsection{Country association additional metric}
\label{metric}

The metric that use the TFIDF metric assign to each user $u$ a country using the following procedure:
\begin{enumerate}
	\item  $\text{TFIDF}_{IP,u} = {tf_{IP,u}} \times log\left( {\frac{N}{{{df_{IP}}}}} \right) $, where:
  \begin{itemize}
  \item $tf_{IP,u}$ is the number of occurrences of that $IP$ in user $u$.
  \item $df_{IP}$ is the number of users that have used that $IP$.
  \item $N$ is total number of users.
  \end{itemize}

  \item We choose as user $u$'s country the most present country among the top $\text{TFIDF}_{IP_i,u}$ measures for the user. We consider as top those values that cover 50\% of the cumulative sum of the TFIDF values for the user.

\end{enumerate}
\smallskip

\begin{equation}
\label{eq:1}
{C_i} = \mathop {\arg \max }\limits_{c \in {\rm{Contries}}} \left\{ {\mathop {{\rm{median}}}\limits_{IP \in I{P_{i,c}}} \left( {{\rm{Ratio}}\left( {IP} \right)} \right) \cdot \frac{{\# I{P_{i,c}}}}{{\# I{P_i}}}} \right\}
\end{equation}
$IP_{i,c}$ = IP addresses that belongs to country $c$ and it is used by user $i$\\
$\#IP_{i,c}$ = number of time user $i$ use $IP_{i,c}$\\
$\#IP_i$ = total number of IP usage from user $i$\\
$Ratio\left( {IP} \right) = \frac{{\# I{P_{i,c}}}}{{\# IP}}$ with $\#IP$ as overall IP usage in the Bitcoin network\\

\end{document}